\documentclass[a4paper,12pt,twoside]{scrartcl}
\usepackage[latin1]{inputenc}
\usepackage[english]{babel}
\usepackage[margin=3cm]{geometry}
\usepackage{amsmath}
\usepackage{amssymb}
\usepackage{graphicx}
\usepackage{fancyhdr}

\newcommand{\chTrack}[1]{#1}
\newcommand{\espu}[1]{}
\newcommand{\aggiu}[1]{#1}
\usepackage[hang,small,bf]{caption}

\usepackage{color}
\newcommand{\mca}[1]{{\mathcal #1}}
\let\embib\normalfont   
\usepackage{tabulary}

\makeatletter
\def\@biblabel#1{\@ifnotempty{#1}{#1.}}
\makeatother

\begin{document}
\thispagestyle{empty}
\begin{center}\sffamily\mdseries\upshape
\phantom{malaguti}
 \vspace{0.5cm}
{\bfseries\Huge Predicting epidemic risk\\from past temporal contact data}\\
\end{center}
\begin{center}
 \vspace{1cm}
Eugenio Valdano\textsuperscript{a,b}, 
Chiara Poletto\textsuperscript{a,b},
Armando Giovannini\textsuperscript{c},
Diana Palma\textsuperscript{c},
Lara~Savini\textsuperscript{c},
and Vittoria Colizza\textsuperscript{a,b,d}\\
\end{center}
\vspace{0.5cm}
\begin{flushleft}
{\footnotesize
\textsuperscript{a}INSERM, UMR-S 1136, Institut Pierre Louis d'Epid\'emiologie et de Sant\'e Publique, F-75013\\
27 rue Chaligny, 75571 Paris, France.\\
\textsuperscript{b}Sorbonne Universit\'es, UPMC Univ Paris 06, UMR-S 1136, Institut Pierre Louis d'Epid\'emiologie et de Sant\'e Publique, F-75013\\
 27 rue Chaligny, 75571 Paris, France.\\
\textsuperscript{c}Istituto Zooprofilattico Sperimentale Abruzzo-Molise G. Caporale\\
 Campo Boario, 64100 Teramo, Italy.\\
\textsuperscript{d}ISI Foundation\\
 Via Alassio 11/c, 10126 Torino, Italy.\\
}
\end{flushleft}

\newpage
\thispagestyle{empty}

{\sffamily\footnotesize
\centering{\bfseries Abstract}\\
 Understanding how epidemics spread in a system is a crucial step to prevent and control outbreaks, with broad implications on the system's functioning, health, and associated costs.  This can be achieved by identifying the elements at higher risk of infection and implementing targeted surveillance and control measures. One important ingredient to consider is the pattern of disease-transmission contacts among the elements, however lack of data or delays in providing updated records may hinder its use, especially for time-varying patterns. Here we explore to what extent it is possible to use past temporal data of a system's pattern of contacts to predict the risk of infection of its elements during an emerging outbreak, in absence of updated data. We focus on two real-world temporal systems; a livestock displacements trade network among animal holdings, and a network of sexual encounters in high-end prostitution.  We define the node's loyalty as a local measure of its tendency to maintain contacts with the same elements over time, and uncover important non-trivial correlations with the node's epidemic risk. We show that a risk assessment analysis incorporating this knowledge and based on past structural and temporal pattern properties provides accurate predictions for both systems. Its generalizability is tested by introducing a theoretical model for generating synthetic temporal networks. High accuracy of our predictions is recovered across different settings, while the amount of possible predictions is system-specific. The proposed method can provide crucial information for the setup  of targeted intervention strategies.
}

\newpage

\section*{Introduction}

Being able to promptly identify who, in a system, is at risk of infection during an outbreak is key to the efficient control of the epidemic. The explicit pattern of potential disease-transmission contacts  has been extensively used to this purpose in the framework of theoretical studies of epidemic processes, uncovering the role of the pattern's properties in the disease propagation and epidemic outcomes~\cite{pastorsatorras2001,lloyd2001,newman2002,cohen2003,keeling2005,riley2007,colizza2006,brockmann2013}. These studies are generally based on the assumption that the entire pattern of contacts can be mapped out or that its main properties are known. Although such knowledge would
be a critical requirement to conduct risk assessment analyses in real-time, which need to be based on the updated and accurate description of the contacts relevant to the outbreak under study~\cite{riley2003}, it can hardly be obtained in reality. Given the lack of such data,  analyses generally refer to the most recent available knowledge of contact data, implicitly assuming a non-evolving pattern. 

The recent availability of time-resolved data characterizing connectivity patterns in various contexts~\cite{eckmann2004,onnela2007,rybski2009, gautreau2009,vernon2009,cattuto2010,rocha2010,salathe2010a,tang2010,bajardi2011a, miritello2011,karsai2012, holme2012} has inevitably weakened the non-evolving assumption, bringing new challenges to the assessment of nodes' epidemic risk. Traditional centrality measures used to identify vulnerable elements or influential spreaders for epidemics circulating on static networks~\cite{pastorsatorras2001,lloyd2001,cohen2003,albert2000,cohen2001,freeman1979,friedkin1991,holme2004,kitsak2010,salathe2010b,natale2011} are unable to provide meaningful information for their control, as these quantities strongly fluctuate in time once computed on the evolving networks~\cite{bajardi2011a,rocha2011}. An element of the system may thus act as {\itshape superspreader} in a past configuration of the contact network, having the ability to potentially infect a disproportionally larger amount of secondary contacts than other elements~\cite{galvani2005}, and then assume a more peripheral role in the current pattern of contact or even become isolated from the rest of the system~\cite{bajardi2011a}. 
If the rules driving the change of these patterns over time are not known, what information can be extracted from past contact data to infer the  risk of infection for an epidemic unfolding on the current (unknown) pattern?  

Few studies have so far tried to answer this question by exploiting temporal information to control an epidemic through targeted immunization. They are based on the extension to temporal networks~\cite{lee2012, starnini2013} of the so-called acquaintance immunization protocol~\cite{cohen2003} introduced in the framework of static networks that prescribes to vaccinate a random contact of a randomly chosen element of the system. In the case of contacts relevant for the spread of sexually transmitted infections, Lee et al. showed that the most efficient protocol consists in sampling elements at random and vaccinating their latest contacts~\cite{lee2012}. The strategy is based on local information gathered from the observation and analysis of past temporal data, and it outperforms static-network protocols. Similar results are obtained for the study of face-to-face contact networks relevant for the transmission of acute respiratory infections in a confined setting, showing in addition that a finite amount of past network data is in fact needed to devise efficient immunization protocols~\cite{starnini2013}.
 
The aim of these studies is to provide general protocols of immunization over all possible epidemiological conditions of the disease (or class of diseases) under study. For this reason,  protocols are tested through numerical simulations and results are averaged over  starting seeds and times to compare their performance. Previous work has however shown that epidemic outcomes may strongly depend on the temporal and geographical initial seed of the epidemic~\cite{bajardi2012}, under conditions of large dynamical variability of the network and  absence of stable structural backbones~\cite{bajardi2011a}. Our aim is therefore to focus on a specific epidemiological condition relative to a given emerging outbreak in the population, resembling a realistic situation of public health emergency. We focus on the outbreak initial phase prior to interventions when facing the difficulty that some infected elements in the population are not yet observed. The objective is to assess the risk of infection of nodes to inform targeted surveillance, quarantine and immunization programs, assuming the lack of knowledge of the explicit contact pattern on which the outbreak is unfolding. Knowledge is instead gathered from the analysis of the full topological and temporal pattern of past data (similarly to previous works~\cite{lee2012,starnini2013}), coupled, in addition, with epidemic spreading simulations performed on such data under the same epidemiological conditions of the outbreak under study.
More specifically, we propose an egocentric view of the system and assess whether and to what extent the node's tendency of repeating already established contacts is correlated with its probability of being reached by the infection.
Findings obtained on past available contact data are then used to predict the infection risk in the current unknown epidemic situation. We apply this risk assessment analysis to two large-scale empirical datasets of temporal contact networks~--~cattle displacements between premises in Italy~\cite{bajardi2011a,natale2009}, and sexual contacts in high-end prostitution~\cite{rocha2010}~--~and evaluate its performance through epidemic spreading simulations. 
We also introduce a  model to generate synthetic time-varying networks retaining the basic mechanisms observed in the empirical networks considered, in order to explain the results obtained by the proposed risk assessment strategy within a general theoretical framework.

\section*{Results and Discussion}
The cattle trade network is extracted from the complete dataset reporting on time-resolved bovine displacements among animal holdings in Italy~\cite{bajardi2011a,natale2009} for the period 2006--2010, and it represents the time-varying contact pattern among the $215,264$ premises composing the system. The sexual contact network represents the connectivity pattern of sexual encounters extracted from a Web-based Brazilian community where sex buyers provide time-stamped rating and comments on their experiences with escorts~\cite{rocha2010}. 

The five-years data of the livestock trade network show that stationary properties at the global level co-exist with an active non-trivial local dynamics. The probability distributions of several quantities measured on the different yearly networks are considerably stable over time, as e.g. shown by the in-degree distribution reported in \chTrack{Fig~\ref{fig:datasets_analysis}A}, where the in-degree of a farm measures the number of premises selling cattle to that farm. 
These features, however, result from highly fluctuating underlying patterns of contacts,  never preserving more than 50\% of the links from one yearly configuration to another (\chTrack{Fig~\ref{fig:datasets_analysis}C}), notwithstanding the seasonal annual pattern due to repeating cycles of livestock activities~\cite{robinson2007,kao2006} (see Supplementary Material).
Similar findings are also obtained for the sexual contact network (\chTrack{Fig~\ref{fig:datasets_analysis}B-D}), where the lack of an intrinsic cycle of activity characterizing the system leads to smaller values of the overlap between different configurations ($<10\%$). In this case we consider semi-annual configurations, an arbitrary choice that allows us to extract six network configurations in a timeframe exhibiting an approximately stationary average temporal profile of the system, 
after discarding an initial transient time period from the data~\cite{rocha2010}. Different time-aggregating windows are also considered (see the Materials and Methods section and Supplementary Material for additional details).

\subsection*{Loyalty}
The observed values of the overlap of the time-resolved contact networks in terms of the number of links preserved are a measure of the degree of memory contained in the system. This is the outcome of the temporal activity of the elements of the system that reshape up to 50\% or 90\% of the contacts of the network (in the cattle trade case and in the sexual contact case, respectively), through nodes' appearance and disappearance, and neighborhood restructuring. 
By framing the problem in an egocentric perspective, we can explore the behavior of each single node of the system in terms of its tendency to remain active in the system and re-establish connections with the same partners vs. the possibility to change partners or make no contacts. We quantitatively characterize this tendency by introducing the {\itshape loyalty} $\theta$, a quantity that measures the fraction of preserved neighbors of a node for a pair of two consecutive network configurations in time, $c-1$ and $c$. If we define $\mca{V}_i^{c-1}$ as the set of neighbors of node $i$ in configuration $c-1$, then  $\theta_i^{c-1,c}$ is given by the Jaccard index between $\mca{V}_i^{c-1}$ and $\mca{V}_i^{c}$:
\begin{equation}
\theta_i^{c-1,c} = \frac{ \left| \mca{V}_i^{c-1} \cap \mca{V}_i^{c} \right| }{ \left| \mca{V}_i^{c-1} \cup \mca{V}_i^{c} \right| }\,.
\label{eq:loyalty}
\end{equation}
Loyalty takes values in the interval $\left[0,1\right]$, with $\theta=0$ indicating that no neighbors are retained, and $\theta=1$ that exactly the same set of neighbors is preserved ($\mca{V}_i^{c-1} = \mca{V}_i^{c}$). It is defined for discrete time windows ($c,c+1$) and in general it depends on the aggregation interval chosen to build network configurations.

In case the network is directed, as for example the cattle trade network, $\theta$ can be equivalently computed on the set $\mca{V}_{in,i}^c$ of incoming contacts or on the set of neighbors of outgoing connections, $\mca{V}_{out,i}^c$, depending on the system-specific interpretation of the direction and on the interest in one phenomenon or the opposite.  This measure originally finds its inspiration in the study of livestock trade networks, where a directed connection from holding $A$ to holding $B$ indicates that $B$ purchased a livestock batch from $A$, which  was then displaced along the link direction $A \to B$. If we compute $\theta$ on the incoming contacts of the cattle trade network, we thus quantify the propensity of each farmer to repeat business deals with the same partners when they purchase their cattle. This concept is at the basis of many loyalty or fidelity programs that propose explicit marketing efforts to incentivize the reinforcement of loyal buying behavior between a purchasing client and a selling company~\cite{sharp1997}, and corresponds to a principle of exclusivity in selecting economic and social exchange partners~\cite{podolny1994,sorenson2006}. Analogously, in the case of the sexual contact network we consider the point of view of sex buyers. 
Formally, our methodology can be carried out with the opposite point of view, by considering out-degrees with  loyalties being computed on out-neighbors. Our choice is arbitrary and inspired by the trade mechanism underlying the network evolution.

Other definitions of similarity to measure the loyal behavior of a node are also possible. In Supplementary Material we compare and discuss alternative choices.
\chTrack{For the sake of clarity all symbols and variables used in the article are reported in Table~\ref{tab:variables}.}
Finally, other mechanisms different from fidelity strategies may be at play that result in the observed behavior of a given node. In absence of additional  knowledge on the behavior underlying the network evolution, we focus on the loyalty $\theta$ to explore whether it can be used as a possible indicator for infection risk, as illustrated in the following subsection.

The distributions of loyalty values, though of different shapes across the two datasets, display no considerable variation moving along consecutive pairs of configurations of each dataset (\chTrack{Fig~\ref{fig:loyalty}A-B} and Supplementary Material), once again indicating the overall global stability of  system's properties in time and confirming the results observed for the degree. A diverse range of behaviors in establishing new connections vs. repeating existing ones is observed, similarly to the stable or exploratory strategies found in human communication~\cite{miritello2013}. Two pronounced peaks are observed for $\theta=0$ and $\theta=1$, both dominated by low degree nodes for which few loyalty values are allowed, given the definition of Eq.~(\ref{eq:loyalty}) (see Supplementary Material for the dependence of $\theta$ on nodes' degree and its analytical understanding). The exact preservation of the neighborhood structure ($\theta=1$) is more probable in the cattle trade network than in the sexual contact network ($P(\theta=1)$ being one order of magnitude larger), in agreement with the findings of a higher system-wide memory reported in \chTrack{Fig~\ref{fig:datasets_analysis}}. Moreover, the cattle trade network exhibits the presence of high loyalty values (in the range $\theta \in [0.7,0.9]$), differently from the sexual contact network where $P(\theta)$ is always equal to zero in that range except for one pair of consecutive configurations giving a positive probability for $\theta=0.8$. Farmers in the cattle trade network thus display a more loyal behavior in purchasing cattle batches from other farmers with respect to how sex buyers establish their sexual encounters in the analyzed sexual contact dataset. 

For the sake of simplification, we divide the set of nodes composing each system into the subset of {\itshape loyal nodes} having $\theta$ greater than a given threshold $\epsilon$, and the subset of {\itshape disloyal nodes} if instead $\theta<\epsilon$. We call hereafter these classes as {\itshape loyalty statuses L} and {\itshape D}, respectively, and we will later discuss the role of the chosen value for $\epsilon$.

\subsection*{Epidemic simulations and risk of infection}
Both networks under study represent substrates offering potential opportunities for a pathogen to diffuse in the corresponding populations. 
Sexually transmitted infections spread among the population of individuals through sexual contacts~\cite{morris1997,anderson1991},
whereas livestock infectious diseases (e.g. Foot-and-mouth disease~\cite{keeling2001}, Bluetongue virus~\cite{saegerman2008}, or BVD \cite{tinsley2012}) can be transmitted from farm to farm mediated by the movements of infected animals (and vectors, where relevant), potentially leading to a rapid propagation of the disease on large geographical scales. 

As a model for disease-transmission  on the network of contacts we consider a discrete-time Susceptible-Infectious compartmental approach~\cite{anderson92}.
No additional details characterizing the course of infection are considered here (e.g. recovery dynamics), as we focus on a simplified theoretical picture of the main mechanisms of pathogen diffusion and their interplay with the network topology and time-variation, for the prediction of the risk of infection. The aim is to provide a general and conceptually simple framework, leaving to future studies the investigation of more detailed and realistic disease natural histories.

At each time step, an infectious node can transmit the disease along its outgoing links to its neighboring susceptible nodes that become infected and can then propagate the disease further in the network. Here, we consider a deterministic process for which the contagion occurs with probability equal to 1, as long as there exist a  link connecting the infectious node to a susceptible one. Although a crude assumption, this allows us to simplify the computational aspects while focusing on the risk prediction. The corresponding stochastic cases exploring lower probabilities of transmission per link are reported in Supplementary Material.

We focus on the early phase of the spreading simulations, defined as the set of nodes infected up to simulation time step $\tau=6$. This choice allows us to study invasion stage only, while the epidemic is no more trivially confined to the microscopic level. Additional choices for $\tau$ have been investigated showing that they do not alter our findings (see Supplementary Material). Network configurations are kept constant  during  outbreaks, assuming diseases spread faster than network evolution, at least during their invasion stage. Examples of incidence curves obtained by the simulations are reported in Supplementary Material.

Livestock disease spread is often modeled by assuming that premises are the single discrete units of the spreading processes and neglecting the possible impact of within-farm dynamics~\cite{keeling2005b}. This is generally considered in the study of highly contagious and rapid infections, and corresponds to regarding a farm as being infected as soon as it receives the infection from neighboring farms following the transport of contagious animals. Under this assumption, both case studies can be analyzed in terms of networks of contacts for disease transmission. 
In addition, for sake of simplicity, we do not take into account the natural definition of link weights on cattle network, representing  the size of the moved batches. In Supplementary Material we generalize our methodology to the weighted case, including a weighted definition of loyalty, reaching results similar to the unweighted case.

We consider an emerging epidemic unfolding on a network configuration $c$ and starting from a single node (seed $s$), where the rest of the population of nodes is assumed to be initially susceptible. The details on the simulations are reported in the Material and Methods section. We define $\mca{I}_s^c$  the set of nodes infected during the early stage invasion. In order to explore how the network topology evolution alters the spread of the disease, we consider an outbreak unfolding on the previous configuration of the system, $c-1$, and characterized by the same epidemiological conditions (same epidemic parameters and same initial seed $s$). By comparing the set of infected nodes $\mca{I}_s^{c-1}$ obtained in configuration $c-1$ to $\mca{I}_s^c$, we can assess changes in the two sets and how these depend on the nodes' loyalty. We define a node's infection potential $\pi_L^{c-1,c}(s)$ ($\pi_D^{c-1,c}(s)$)  measuring the probability that a node will be infected in configuration $c$ by an epidemic starting from seed $s$, given that it was infected in configuration $c-1$ under the same epidemiological conditions and provided that its loyalty status is $L$ ($D$): 
\begin{align*}
 & \pi_L^{c-1,c}(s) \stackrel{def}{=} \mbox{Prob}\left[    i\in \mca{I}_s^{c} \; | \;  i\in \mca{I}_s^{c-1} \mbox{ and } i\in \{L\}  \right], \\
 & \pi_D^{c-1,c}(s) \stackrel{def}{=} \mbox{Prob}\left[    i\in \mca{I}_s^{c} \; | \;  i\in \mca{I}_s^{c-1} \mbox{ and } i\in \{D\}  \right], \\
\end{align*}
where $i$ is a node of the system.
$\pi_L$ and $\pi_D$ thus quantify the effect of the temporal stability of the network at the local level (loyalty of a node) on the stability of a macroscopic process unfolding on the network (infection). They depend on the seed chosen for the start of the epidemic, on the pair $(c-1,c)$ of network configurations considered along its evolution, and also on the threshold value $\epsilon$ assumed for the definition of the loyalty status of the nodes. 

By exploring all seeds and computing the infection potentials for different couples of years, we obtain sharply peaked probability distributions of $\pi_L$ and $\pi_D$  around values that are well separated along the $\pi$ axis. Results are qualitatively similar in both cases under study, with peaks reached for $\pi_L/\pi_D\simeq 2.5$ in the cattle trade network and $\pi_L/\pi_D\simeq 3$ in the sexual trade network (\chTrack{Fig~\ref{fig:pi_T}A-B}). An observed  infection in $c-1$, based on the knowledge of the epidemiological conditions and no information on the network evolution, is an indicator of an infection risk for the same epidemic in $c$ more than twice larger for loyal farms with respect to disloyal farms. Analogously, loyal sex buyers have a threefold increase in their infection potential with respect to individuals having a larger turnover of partners. Remarkably, small values of loyalty threshold $\epsilon$ are able to correctly characterize the loyal behavior of nodes with status $L$. Results shown in \chTrack{Fig~\ref{fig:pi_T}A-B} are obtained for $\epsilon=0.1$. Findings are however robust against changes in the choice of the threshold value, as this is induced by the peculiar bimodal shape of the probability distribution curves for the loyalty (see Supplementary Material). This means that intermediate values of the local stability of the nodes (i.e. $\theta>\epsilon$) imply that a possible risk of being infected is strongly stable, regardless of the dynamics of the network evolution. Valid for all possible seeds and epidemiological conditions, this result indicates that the loyalty of a node can be used as an indicator for the node's risk of infection, which has important implication for the spreading predictability in case an outbreak emerges. 

These results are obtained for temporally evolving networks where no further change induced by the epidemic is assumed to occur. Focusing on the initial stage of the outbreak, we disregard the effect of interventions (e.g. social distancing, quarantine of infectious nodes, movements bans) or of adaptive behavior following awareness~\cite{robinson2007,gross2006, funk2010, shaw2010, meloni2011, bajardi2011b}. Such assumption relies on the study's focus on the initial stage of the epidemic that may be characterized by a silent spreading phase with propagation occurring before the alert or outbreak detection takes place; or, following an alert, by a contingent delay in the implementation of intervention measures.

\subsection*{Risk assessment analysis}
The observed relationship between loyalty and infection potential can be used to define a strategy for the risk assessment analysis of an epidemic unfolding on an unknown networked system at present time, for which we have however information on its past configurations. This may become very useful in practice even in the case of complete datasets, as for example with emerging outbreaks of livestock infectious diseases. Data on livestock movements are routinely collected following European regulations~\cite{EC-livestock}, however they may not be readily available in a real-time fashion upon an emergency, and a certain delay may thus be expected. Following an alert for an emerging livestock disease epidemic, knowledge of past network configurations may instead be promptly used in order to characterize the loyalty of farmers, simulate the spread of the disease on past configurations and thus provide the expected risk of infection for the farms under the ongoing outbreak. The general scheme of the strategy for the risk assessment analysis is composed of the following steps, assuming that the past network configurations $\{c-n,\ldots,c-1,c \}$ are known and that the epidemic unfolds on the unknown configuration $c+1$:
\begin{enumerate}
\item identify the seed $s$ of the ongoing epidemic;
\item characterize the loyalty of the nodes from past configurations by computing $\theta_i^{c-1,c}$ from Eq.~(\ref{eq:loyalty});
\item predict the loyalty of the nodes for the following unknown configuration $c+1$: $\theta_i^{c,c+1}$;
\item simulate the spread of the epidemic on the past configuration $c$ under the same epidemiological conditions of the ongoing outbreak and identify the infected nodes $\mca{I}_s^c$;
\item compute the node epidemic risk for nodes in statuses $L$ and $D$.
\end{enumerate} 
This strategy enables the assessment of the present infection risk (i.e. on configuration $c+1$) for all nodes hit by the simulated epidemic spreading on past configuration $c$ ($\mca{I}_s^c$), not knowing their present pattern of contacts.
It is based on configurations from $c-n$ to $c$ as they are all used to build the probability distributions needed to train our approach. In the cases under study such distributions are quite stable over time so that a small set of configurations ($\{c-2,c-1,c \}$) was shown to be  enough.

To make the above strategy operational, we still need to determine  how we can exploit past data to  predict the evolution of the loyalty of a node in future configurations (step $3$) and use this information to compute nodes epidemic risk (point 5). As with all other variables characterizing the system, indeed, also $\theta$ may fluctuate from a pair of configurations $(c-1,c)$ to another, as nodes may alter their loyal behavior over time, increasing or decreasing the memory of the system across time. Without any additional knowledge or prior assumption on the dynamics driving the system, we measure from available past data the probabilities of (dis)loyal nodes staying (dis)loyal across consecutive configurations, or conversely, of changing their loyalty status. This property can be quantified in terms of probabilities of transition across loyalty statuses. We thus define  $T^c_{LL}(k)$ as the probability that a node with degree $k$ being loyal between configurations $c-1$ and $c$ will stay loyal one step after ($c,c+1$). It is important to note the explicit dependence on the degree $k$ of the node (here defined at time $c$), which may increase or decrease following neighborhood reshaping (it may also assume the value $k=0$  if the node becomes  inactive in  configuration $c$). Analogously, $T^c_{DD}(k)$ is the probability of remaining disloyal. The other two possible transition probabilities are easily obtained as $T_{LD}=1-T_{LL}$ and $T_{DL}=1-T_{DD}$.

\chTrack{Fig~\ref{fig:pi_T}C-D} show the transition probabilities of maintaining the same loyalty status calculated on the two empirical networks for $\epsilon=0.1$. 
Stability in time and non-trivial dependences on the degree of the node are found for both networks. In the cattle trade network, loyal farmers tend to remain loyal with a rather high probability ($T_{LL}>0.6$ for all $k_{in}$ values). In addition, this probability markedly increases with the degree, reaching $T_{LL}\simeq 1$ for the largest values of $k_{in}$. Interestingly, the probability that a disloyal farmer stays disloyal the following year dramatically decreases with the degree, reaching $0$ in the limit of large degree. Among the farmers who purchase cattle batches from a large number of different premises, loyal ones have an increased chance to establish business deals with the same partners the following year, whereas previously disloyal ones will more likely turn to being loyal. 

A similar qualitative dependence on the degree is also found in the sexual contact network, however in this case the probability of remaining disloyal is always very high ($T_{DD} > 0.7$) even for high degrees. $T_{LL}$ shows a relatively more pronounced dependence on $k$, ranging from $0.3$ (low degree nodes) to $0.6$ (high degree nodes). Differently from the farmers behavior, sex buyers display a large tendency to keep a high rate of partners turnover across time. Moreover, the largest probability of preserving sexual partners is obtained when the number of partners is rather large. 

Remarkably, in both networks, transition probabilities are found to be stable across time and are well described by logarithmic functions (with parameters  depending on the system and on $\epsilon$) that can be used to predict the loyalty of nodes in configuration $c+1$ from past data (\chTrack{Fig~\ref{fig:pi_T}C-D}). 
With this information, it is then possible to compute the 
epidemic risk of a node $i$ in configuration $c+1$, having degree $k=k_i^c$ in configuration $c$ and known loyalty status $\{L,D\}$ between configurations $c-1$ and $c$ as follows:
\begin{equation}
\begin{cases}
 \mbox{if loyalty class}=D: \,\,\, \, \rho_i^{c+1} = \pi_{D}^{c,c+1}(s) T_{DD}(k) + \pi_{L}^{c,c+1}(s) T_{DL}(k) \,; \\
 \mbox{if loyalty class}=L: \,\,\, \, \rho_i^{c+1} = \pi_{D}^{c,c+1}(s) T_{LD}(k) + \pi_{L}^{c,c+1}(s) T_{LL}(k) \,.
\end{cases}
\label{eq:epidemic_risk}
\end{equation}

It is important to note that in our framework the epidemic risk is a node property, and not a global characteristic of a specific disease.

\subsection*{Validation}
To validate our strategy of risk assessment, we test our predictions based on past data for the risk of being infected in configuration $c+1$ on the results of an epidemic simulation explicitly performed on the supposedly unknown configuration $c+1$. We consider the set of nodes $\mca{I}_s^{c}$ for which we are able to provide risk predictions and divide it into two subsets, according to their predicted risk of infection $\rho_i^{c+1}$. We indicate with $\mca{I}_{s,h}^{c}$ the top $25\%$ highest ranking nodes, and with $\mca{I}_{s,l}^{c}$ all the remaining others. We then compute the fraction $P_h$ of nodes in the subset $\mca{I}_{s,h}^{c}$, i.e. predicted at high risk, that belong to the set of infected nodes $\mca{I}_s^{c+1}$ in the simulated epidemic aimed at validation. Analogously, $P_l$ measures the fraction of nodes in $\mca{I}_{s,l}^{c}$ that are reached by the infection in the simulation on $c+1$. In other words, $P_h$ ($P_l$) represents the probability for a node having a high (low) risk of infection to indeed get infected. The accuracy of the risk assessment analysis can thus be measured in terms of the relative risk ratio $\nu=P_h/P_l$, where values $\nu \le 1$ indicate negative or no correlation between our risk predictions and the observed infections, whereas values $\nu>1$ indicate that the prediction is informative. For both networks we find a significant correlation, signaled by the distributions of the relative risk ratio $\nu$ peaking around  values $\nu>1$ (\chTrack{Fig~\ref{fig:validation}A-B}).
The peak positions ($\nu\simeq 1.4$ and $\nu\simeq 1.7$ for cattle and sex, respectively) are remarkably close to the benchmark values represented by the distributions computed on the training sets (red lines in \chTrack{Fig~\ref{fig:validation}A-B}).
In addition, the comparison with the distributions from a null model obtained by reshuffling the infection statuses of nodes (dotted curves peaking around $\nu=1$ in \chTrack{Fig~\ref{fig:validation}A-B}) further confirms the accuracy of the approach.
Findings are robust against changes of the value used to define $\mca{I}_{s,h}^{c}$ or against alternative definitions of this quantity (see Supplementary Material).
 
One other important aspect to characterize is the predictive power of our risk assessment analysis. Our predictions indeed are limited to the set $\mca{I}_s^{c}$ of nodes that are reached in the  simulation performed on past data, proxy for the future outbreak. If a node is not infected by the simulation unfolding on configuration $c$ or it is not active at that given time, our strategy is unable to provide a risk assessment for that node in the future. We can then quantify the predictive power $\omega$ as the fraction of infected nodes for which we could provide the epidemic risk, i.e.  $\omega_s^{c,c+1}  = \left| \mca{I}_s^{c+1} \cap \mca{I}_s^{c} \right| / \left| \mca{I}_s^{c+1} \right|$. High values of $\omega$ indicate that few infections are missed by the risk assessment analysis.
\chTrack{Fig~\ref{fig:validation}C-D} display the distributions $P(\omega)$ obtained for the two case studies, showing that a higher predictive power is obtained in the cattle trade network (peak at $\omega\simeq 60\%$) with respect to the sexual contact network (peak at $\omega\simeq 40\%)$.
Our methodology can potentially be applied to a wide range of networks, other than the ones presented here, as shown with the example of human face-to-face proximity networks relevant for the spread of respiratory diseases reported in Supplementary Material.

We also tested whether our risk measure represents a significant improvement in prediction accuracy with respect to simpler and more immediate centrality measures (namely, the degree). Through a multivariate logistic regression, in Supplementary Material we show that our definition of node risk is predictor of infection even after adjusting for node degree.

\subsection*{Memory driven dynamical model}
\begin{figure}[htp!]
 \begin{center}
\includegraphics[width=1\linewidth]{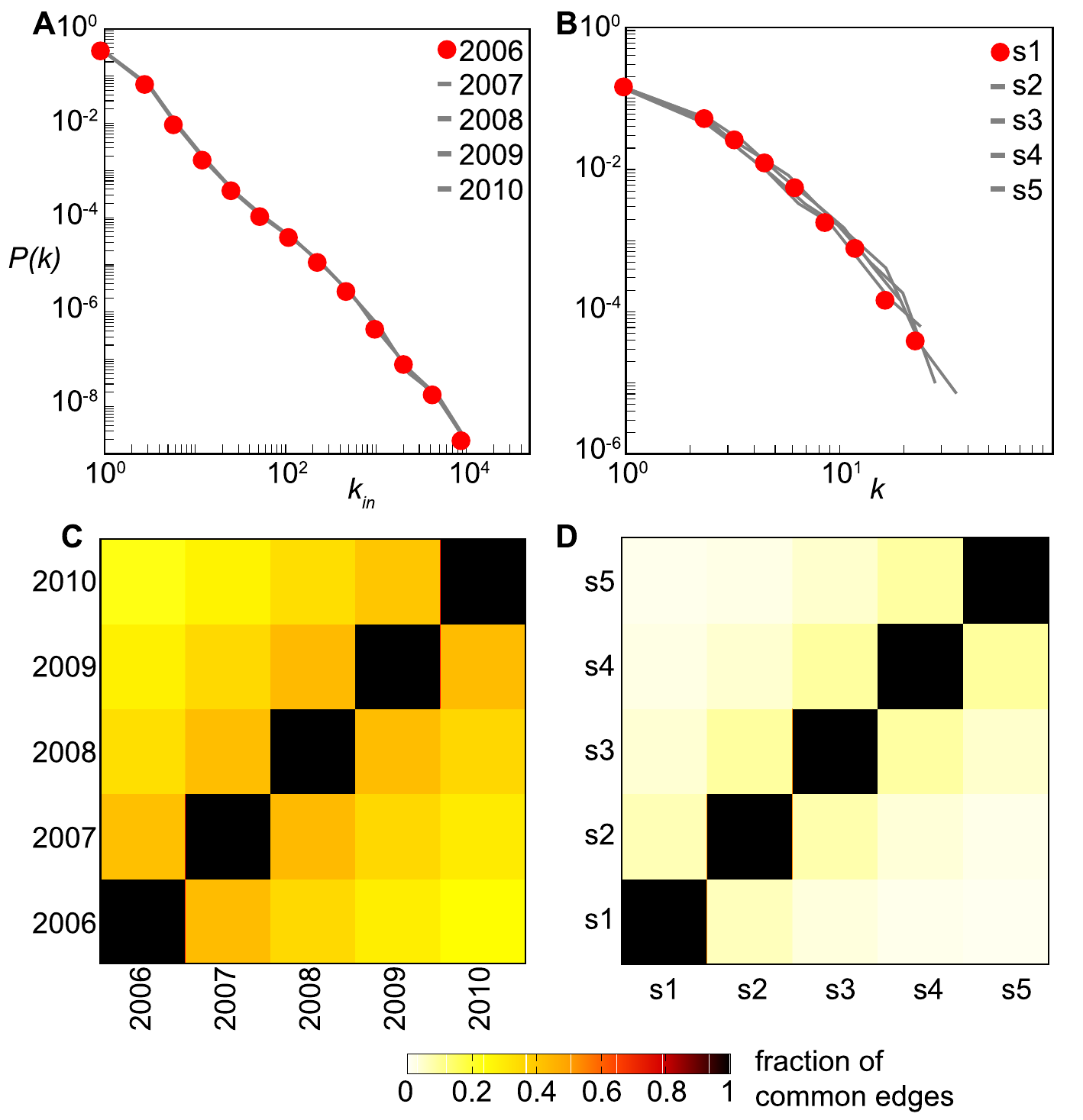}
  \end{center}
 \caption{
{\bfseries Structural and temporal properties of the cattle trade network and of the sexual contact network.}
({\itshape A}), ({\itshape B}): premises in-degree distributions in the cattle trade network and sex customers degree distribution in the sexual contact network, respectively. Distributions for different  configurations of the networks are superimposed in both cases.
 ({\itshape C}), ({\itshape D}): fraction of common edges contained in two  configurations of the network, for the cattle trade network and the sexual contact network, respectively. In ({\itshape B}), ({\itshape D}) $s$ stands for semester, the aggregation interval of each configuration.
 }
 \label{fig:datasets_analysis}
 \end{figure}

\begin{figure}[htp!]
  \begin{center}
\includegraphics[width=0.95\linewidth]{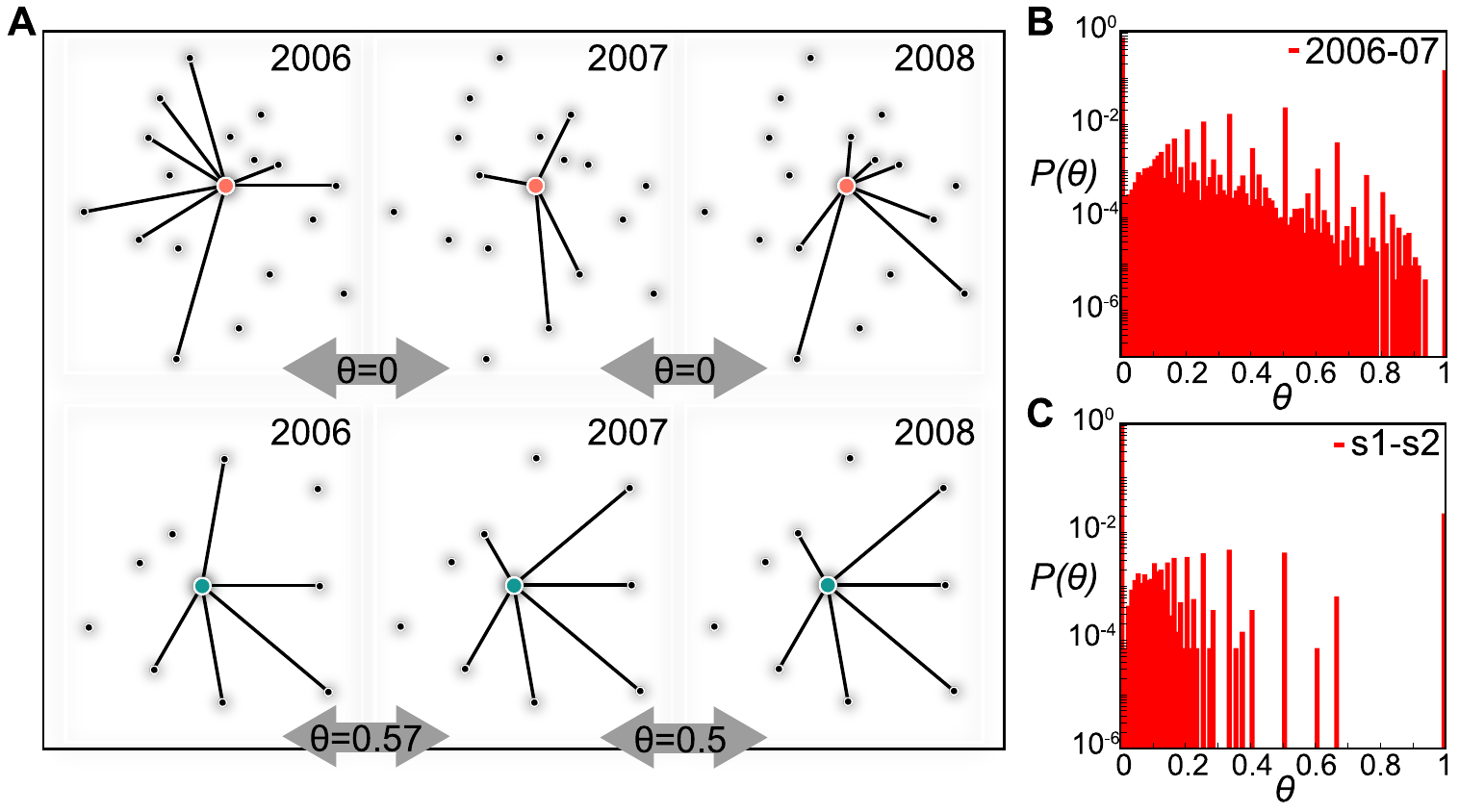}
 \end{center}
 \caption{
 {\bfseries Loyalty.} 
({\itshape A}) Visualization of the neighborhood of two different farms in the cattle trade network (orange node, characterized by low loyalty, and green node, characterized by high loyalty) and corresponding loyalty values computed on three consecutive configurations (2006, 2007, 2008).  ({\itshape B}), ({\itshape C}): Loyalty distributions in the cattle trade network and in the sexual contact network, respectively. Histograms refer to the first pair of consecutive configurations for visualization purposes, all other distributions being reported in Supplementary Material and showing stability across time.
 }
 \label{fig:loyalty}
 \end{figure}

\begin{figure}[htp!]
 \begin{center}
\includegraphics[width=1\linewidth]{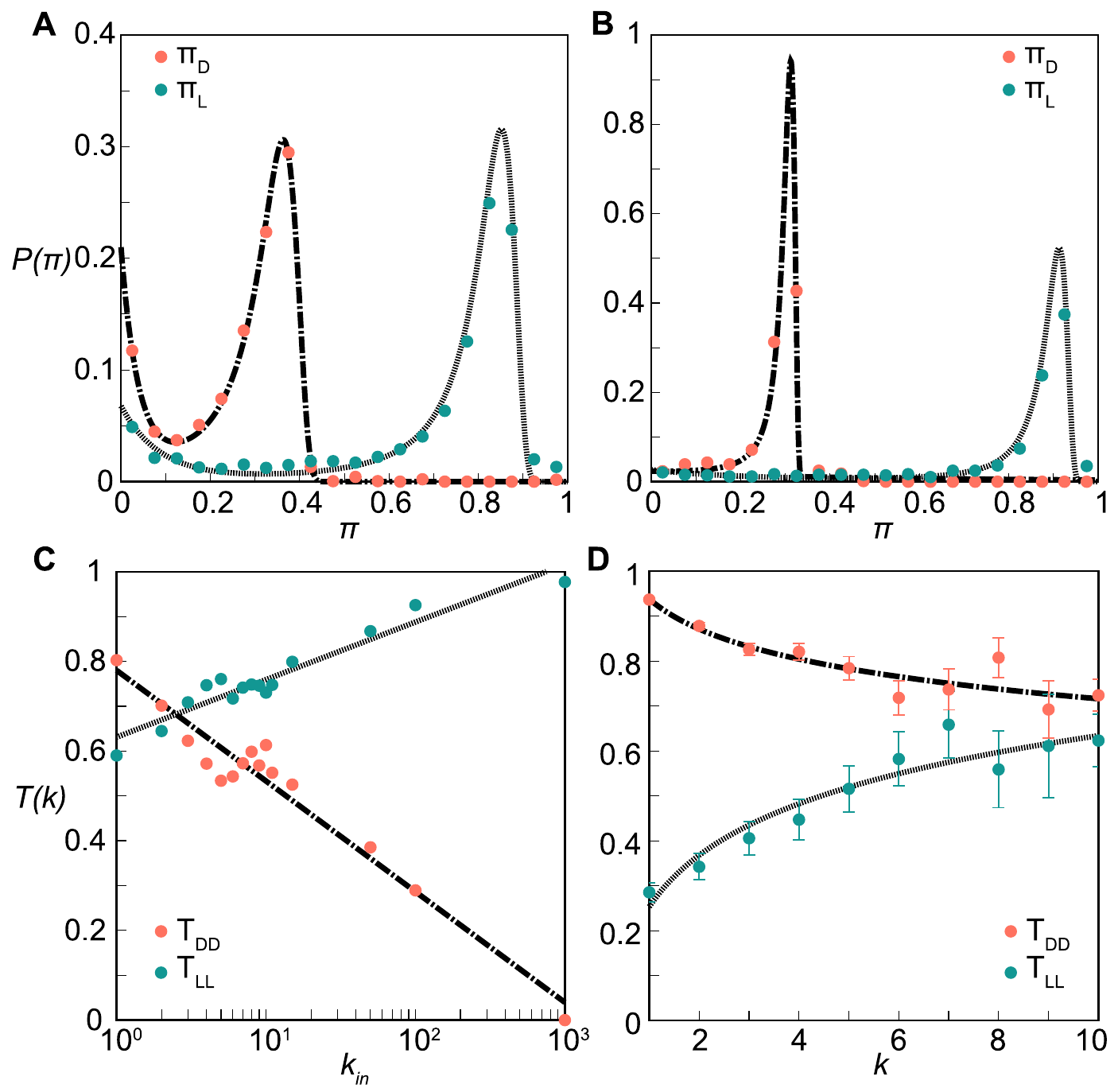}
  \end{center}
 \caption{
  {\bfseries Infection potentials and loyalty transitions.} ({\itshape A}), ({\itshape B}): Probability distributions of the infections potentials for loyal ($\pi_L$, green) and disloyal nodes ($\pi_D$, orange), for the cattle trade network and the sexual contact network, respectively. Loyalty is set with a threshold $\epsilon=0.1$. Dashed lines show the fit with a Landau+exponential model (see Material and Methods). 
 ({\itshape C}), ({\itshape D}): Loyalty transition probabilities between loyal statuses ($T_{LL}(k)$, green) and disloyal statuses ($T_{DD}(k)$, orange) as functions of the degree $k$ of the node, for the cattle trade network and the sexual contact network, respectively. Dashed lines represent the logarithmic models: $T_{DD}(k)=0.78-0.11\log k$, and $T_{LL}(k)=0.63+0.06\log k$ for the cattle trade network; $T_{DD}(k)=0.94-0.10\log k$, and $T_{LL}(k)=0.25+0.17\log k$ for the sexual contact network.
 Transition probabilities are computed as frequencies in the datasets under study. The error bars here represent one binomial standard deviation from these frequencies.
In ({\itshape C}) the error bars are smaller than the size of the points. A single pair of configurations is considered here as example; the behavior observed is the same for all the pair of configurations.
 }
 \label{fig:pi_T}
 \end{figure}

\begin{figure}[htp!]
  \begin{center}
\includegraphics[width=1\linewidth]{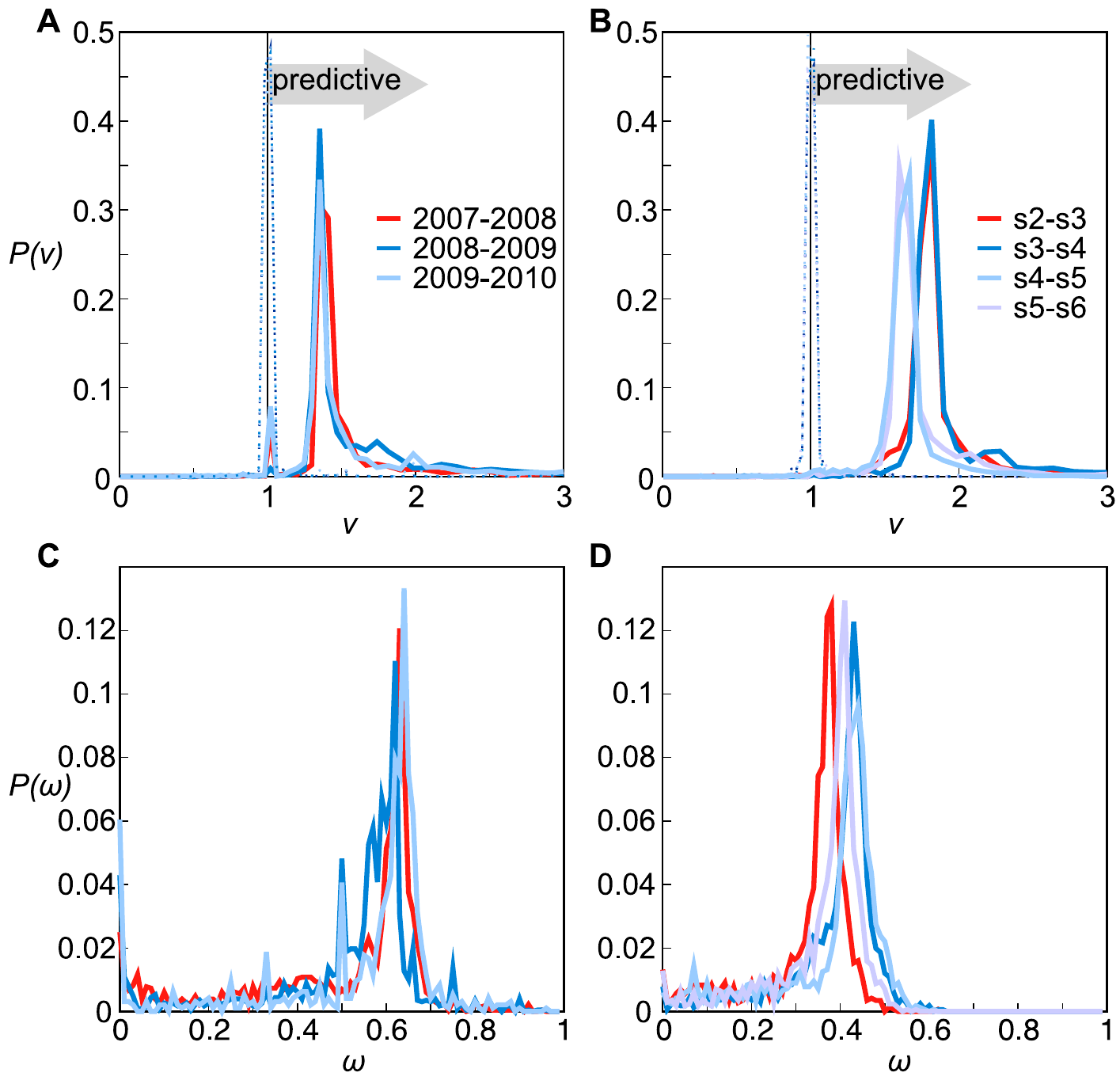}
 \end{center}
 \caption{
 {\bfseries Validation of the risk assessment analysis.}
 ({\itshape A}), ({\itshape B}): Probability distributions of the risk ratio $\nu$ for the cattle trade network and the sexual contact network, respectively. Red lines are computed on training sets (2007-08 for cattle and s2-s3 for sexual contacts). The dashed lines peaking around 1 represent a null model based on reshuffling the infection statuses, i.e. randomly permuting the attribute {\itshape ``actually being infected"} among the nodes for which risk assessment is performed. 
 ({\itshape C}), ({\itshape D}): Probability distributions of the predictive power $\omega$ for the cattle trade network and the sexual contact network, respectively.
}
 \label{fig:validation}
 \end{figure}

The results of the risk assessment analysis obtained from the application of our strategy to the two empirical networks show qualitatively similar results, indicating that the approach is general enough to provide valuable information on the risk of infection in different settings. The observed differences in the predictive power of the approach are expected to be induced by the different temporal behavior of the two systems, resulting in a different amount of memory in preserving links (\chTrack{Fig~\ref{fig:datasets_analysis}}) and different loyalty of  nodes and their time-variations (\chTrack{Fig~\ref{fig:loyalty}} and~\ref{fig:pi_T}\chTrack{C-D}). 

In order to systematically explore the role of these temporal features on the accuracy and predictive power of our approach, we introduce a generic model for the generation of synthetic temporal networks. The model is based on a set of parameters that can be tuned to reproduce the empirically observed features of the two networks, i.e.: {\itshape (i)} the topological heterogeneity of each configuration of the network described by a stable probability distribution (\chTrack{Fig~\ref{fig:datasets_analysis}A-B}); {\itshape (ii)} a vital dynamics to allow for the appearance and disappearance of nodes; {\itshape (iii)} a tunable amount of memory characterizing the time evolution of the network contacts (\chTrack{Fig~\ref{fig:datasets_analysis}C-D}). These specific properties differentiate our approach from the previously introduced models that display instantaneous homogeneous properties for network configurations~\cite{sthele2010,perra2012b,starnini2013b,karsai2014}, reproduce bursty inter-event time distributions but without the explicit introduction of memory~\cite{lee2012, rocha2013,holme2013} or of its control~\cite{starnini2013b}.

\begin{figure}[htp!]
 \begin{center}
\includegraphics[width=1\linewidth]{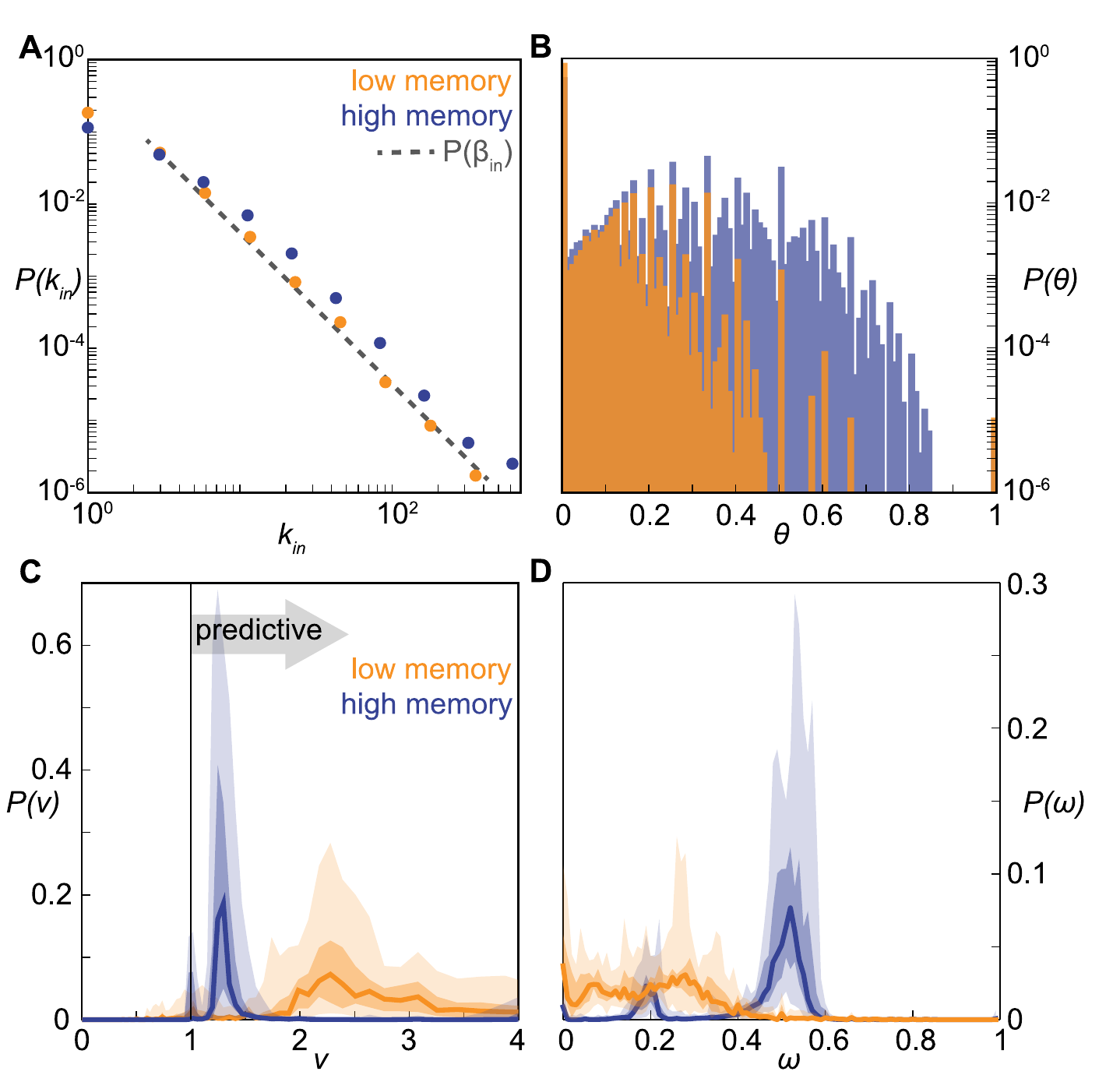}
  \end{center}
 \caption{
 {\bfseries Memory driven dynamical model: model properties and validation of the risk assessment analysis.}
 ({\itshape A}): Probability distributions of  the node in-degree, in the low ($p_\alpha=0.3$) and high memory ($p_\alpha=0.7$) regimes. The slope of the distributions does not depend on $p_\alpha$, and it is forced by the exponent $\gamma$ of the $\beta_{in}$ distribution (dashed line).
 ({\itshape B}): Probability distributions of the loyalty, in the low and high memory regimes.  Distributions are color-coded as in panel ({\itshape a}).
 ({\itshape C}): Probability distributions of the risk ratio $\nu$, in the low and high memory regimes. Lines represent the median values obtained from 50 realizations of the model; darker and lighter shaded areas represent the $50\%$ and $95\%$ confidence intervals.
 ({\itshape D}): Probability distributions of the predictive power $\omega$, in the low and high memory regimes. Medians and confidence intervals are presented as in panel ({\itshape C}). Distributions are color-coded as in panel ({\itshape A}).
 }
\label{fig:model}
 \end{figure}

Based on an iterative network generation approach (see Materials and Methods), we can build an arbitrarily large number of configurations of networks with $10^4$ nodes. They are characterized by stable  in-degree and out-degree heterogeneous distribution across time (\chTrack{Fig~\ref{fig:model}A} where high memory and low memory regimes are displayed) and by  profiles for the probability distribution of the loyalty as in the empirical networks (\chTrack{Fig~\ref{fig:model}B}). The number of nodes with zero loyalty can be computed analytically (see Materials and Methods) and it is confirmed by numerical findings (see Supplementary Material). A high memory regime  corresponds to having nodes in the system that display a highly loyal behavior (e.g., $\theta>0.7$), whereas values in the range $\theta \in\left[0.7,1\right)$ are almost absent in a low memory regime, in agreement with the findings of \chTrack{Fig~\ref{fig:loyalty}}.

Applying the introduced risk assessment analysis to the synthetically generated temporal network, we recover a significant accuracy for both memory regimes (\chTrack{Fig~\ref{fig:model}C}). Different degrees of memory are however responsible for the fraction of the system for which a risk assessment can be made. In networks characterized by higher memory, the distribution of the predictive power $\omega$ has a well defined peak, whereas for lower memory it is roughly  uniform in the range $\omega\in[0,0.4]$ (\chTrack{Fig~\ref{fig:model}D}). Such a regime implies that not enough structure is maintained in the system to control more than $40\%$ of the future infections.  Our risk assessment analysis allows therefore accurate predictions across varying memory regimes characterizing the temporal networks, but the degree of memory impacts the amount of predictions that can be made. The model also shows that the analysis is not affected by the choice of the aggregating time window used to define the network configurations~\cite{holme2013, hoffmann2012, ribeiro2013}, as long as the heterogeneous topological features at the system level and the heterogeneous memory at the node level are kept across aggregation, as observed for the empirical networks under study (see~\cite{bajardi2011a} and Supplementary Material).

\section*{Conclusions}
We introduce a simple measure  to characterize the amount of memory in the time evolution of a networked system. The measure is local and it is empirically motivated from two case studies relevant for disease transmission. By focusing on the degree of loyalty that each node has in establishing connections with the same partners as time evolves, we are able to connect an egocentric view of the system (the node's strategy in establishing its neighborhood over time) to the system's larger-scale properties characterizing the early propagation of an emerging epidemic. 

We uncover a non-trivial correlation between the loyalty of a node and its risk of being infected if an epidemic occurs, given fixed epidemiological conditions, and use this to inform a risk assessment analysis applicable to different settings with no information on the network evolution dynamics.
A theoretical model generating synthetic time-varying networks 
allows us to frame the analysis in a more general perspective and disentangle the role of different features. The accuracy of the proposed risk assessment analysis is stable across variations of the temporal correlations of the system, whereas its predictive power depends on the degree of memory kept in the time evolution. The introduced strategy can be used to inform preventive actions in preparation to an epidemic and for targeted control responses during an outbreak emergency, only relying on past network data.

\section*{Methods}

\subsection*{Datasets}
The cattle trade network is obtained from the database of the Italian national bovine registry recording all cattle displacements due to trade transactions. We consider animal movements during a 5 years time period, from 2006 to 2010, involving 215,264 premises and 2,973,710  directed links. Nodes may be active or inactive depending whether  farms sell/buy cattle in a given timeframe. From the dataset we have removed slaughterhouses ($\sim 1\%$ of the nodes) as they are not relevant for transmission.

The sexual contact network is extracted from an online Brazilian forum where male sex buyers rate and comment on their sexual encounters with female sex sellers~\cite{rocha2010}. Time-stamped posts are used as proxies for sexual intercourse and multiple entries are considered separately, following previous works~\cite{rocha2010,rocha2011}. A total of 13,855 individuals establishing 34,509 distinct sexual contacts are considered in the study, after discarding the initial transient of the community growth~\cite{rocha2010}. Nodes may be active or inactive depending whether individuals use or not the service, and join or quit the community.
Six-months aggregating snapshots are chosen. A different aggregating time window of three months has been tested, obtaining similar results (see Supplementary Material).

\subsection*{Risk of infection}

The distributions of the risk potentials $\pi_L$ and $\pi_D$ reported in \chTrack{Fig~\ref{fig:pi_T}} are modeled with a sum of Landau distribution and an exponential suppression. This family of functions depending on four parameters (see Supplementary Material for the specific functional form) was chosen as it well reproduces  the distribution profiles of the risk potentials, and it was used to compute the nodes' epidemic risk. A goodness of fit was not performed, as this choice was automatically validated in the validation analysis performed on the whole prediction approach.

\subsection*{Memory driven model}
The basic iterative network generation approach allows to build configuration $c+1$  from configuration $c$ through the following steps: 
\begin{itemize}
 \item {vital dynamics}: nodes that are inactive in configuration $c$ become active in $c+1$ with probability $b$, while active nodes become inactive with probability $d$;
 \item {memory}: active nodes maintain same in-neighbors each with probability $p_\alpha$; then they form $\beta_{in}$ new in-stubs, where $\beta_{in}$ is extracted from a power-law  distribution: $P(\beta_{in}) \sim \beta_{in}^{-\gamma}$;
 \item {out-degree heterogeneity}: each node is assigned $\beta_{out}$ out-stubs, where $\beta_{out}$ is drawn from another power-law distribution: $P(\beta_{out}) \sim \beta_{out} ^{-\delta}$. Then each of the in-stubs is randomly matched to an out-stub.
 \end{itemize}

The total set of nodes is considered to be fixed in time, and nodes may be active (i.e. establishing connections) or inactive (i.e. isolated) in a given configuration. All five parameters $b,d,\gamma,p_\alpha,\delta$ are assumed constant in time and throughout the network. The amount of memory in the system is tuned by the interplay of the two parameters $p_{\alpha}$ and $d$. Starting from an arbitrarily chosen initial configuration $c=0$, simulations show that the system rapidly evolves towards a dynamical equilibrium, and successive configurations can be obtained after discarding an initial transient of time. The parameters values used in the paper are: $N=10^4; b=0.7; d=0.2; \gamma = 2.25; \delta = 2.75; p_{\alpha}=0.3,\,0.7$. The influence of such parameters on the network properties is examined in Supplementary Material.

If we denote with $\alpha$ the number of neighbors that a given node keeps across two consecutive configurations $(c-1,c)$, we can express the loyalty simply as:
\begin{equation}
\theta_i^{c-1,c}=\frac{\alpha^{c-1,c}}{\left( k_i^{c} + \beta_{in}^{c}\right)}
\end{equation}
where the superscript $c$ for $\alpha,\beta_{in}$ indicate the values used to build  configuration $c$. The number of nodes with $\theta=0$ as a function of the degree can be computed analytically: {\small$P(\theta_{c,c+1}=0) = d+(1-d)\left(1-p_\alpha\right)^{k_c}$}. Similarly, it is possible to compute the probability $f_{c,c+1}$ that a link present in configuration $c$ is also present in configuration $c+1$. In the Supplementary Material we show that $f_{c,c+1} \simeq (1-d)p_\alpha$ and confirm this result by numerical simulations.


\bibliographystyle{plain}

\newpage

\begin{table}[ht!]
\centering
\begin{tabulary}{\textwidth}{L|L}
 {\bfseries Notation} & {\bfseries Description} \\
 \hline\hline
 $c$ & index for network configurations \\
 \hline
 $\theta$ or $\theta_i^{c-1,c}$ & loyalty of node $i$ between configurations $c-1,c$ \\
 \hline
 $\mca{V}_{in,i}^c,\mca{V}_{out,i}^c$ & set of in(out)-neighbors of $i$ in config $c$ \\
 \hline
 $L$,$D$ & loyalty classes (loyal,disloyal) \\
 \hline
 $\epsilon$ & loyalty threshold \\
 \hline
 $s$ & epidemic seed \\
 \hline
 $\tau$ & duration of the outbreak early stage \\
 \hline
 $\mca{I}_s^c$ & set of infected nodes for outbreak starting from $s$ in config $c$ \\
 \hline
 $\pi_D^{c-1,c}(s),\pi_L^{c-1,c}(s)$ & infection potentials for class $D$ ($L$) computed for seed $s$ between configs $c-1,c$ \\
 \hline
 $k$ & degree (in-degree for the cattle trade network)\\
 \hline
 $T^c_{DD}(k)$, $T^c_{DL}(k)$, $T^c_{LD}(k)$, $T^c_{LL}(k)$ & transition probability from one loyalty class to another \\
 \hline
 $\rho_i^c$ & epidemic risk for node $i$ in config $c$ \\
 \hline
 $\mca{I}_{s,h}^c,\mca{I}_{s,l}^c$ & set of infected nodes with high(low) epidemic risk\\
 \hline
 $P_h,P_l$ & probability of a high(low) risk node to be  infected\\
 \hline
 $\nu = P_h/P_l$ & risk ratio between $P_h,P_l$, measure of accuracy\\
 \hline
 $\omega_s^{c-1,c}$ & predictive power (fraction of infected nodes for which it is possible to compute the epidemic risk)\\
 \hline
 $b,d$ & node probability of becoming active or inactive\\
 \hline
 $p_\alpha$ & node probability of keeping an in-neighbor\\
 \hline
 $\alpha$ & number of kept in-neighbors\\
 \hline
 $\beta_{in}$ & number of new in-neighbors\\
 \hline
 $\beta_{out}$ & number of new out-neighbors\\
 \hline
 $\gamma,\delta$ & exponents of the distributions of $\beta_{in},\beta_{out}$\\
 \hline\hline
\end{tabulary}
\caption{{\bfseries List of variables and their description.} }
\label{tab:variables}
\end{table}

\newpage

\newpage
\thispagestyle{empty}
\fancyhead[LO]{\sffamily Supplementary Material}
\fancyfoot[LE,RO]{\sffamily SM-\thepage}
\setcounter{page}{0}
\phantom{malaguti}
\newpage

  \setcounter{figure}{0}
 \renewcommand\thefigure{S\arabic{figure}}
\renewcommand\thetable{S\arabic{table}}
\appendix


 \thispagestyle{empty}
\begin{center}\sffamily\mdseries\upshape
\phantom{malaguti}
 \vspace{0.5cm}
{\bfseries\Huge Predicting epidemic risk\\from past temporal contact data}\\
\end{center}
\vspace{0.5cm}
\begin{center}\sffamily\mdseries\upshape
 {\Large Supplementary Material}
\end{center}
\begin{center}
 \vspace{1cm}
Eugenio Valdano\textsuperscript{a,b}, 
Chiara Poletto\textsuperscript{a,b},
Armando Giovannini\textsuperscript{c},
Diana Palma\textsuperscript{c},
Lara~Savini\textsuperscript{c},
and Vittoria Colizza\textsuperscript{a,b,d}\\
\end{center}
\vspace{0.5cm}
\begin{flushleft}
{\footnotesize
\textsuperscript{a}INSERM, UMR-S 1136, Institut Pierre Louis d'Epid\'emiologie et de Sant\'e Publique, F-75013\\
27 rue Chaligny, 75571 Paris, France.\\
\textsuperscript{b}Sorbonne Universit\'es, UPMC Univ Paris 06, UMR-S 1136, Institut Pierre Louis d'Epid\'emiologie et de Sant\'e Publique, F-75013\\
 27 rue Chaligny, 75571 Paris, France.\\
\textsuperscript{c}Istituto Zooprofilattico Sperimentale Abruzzo-Molise G. Caporale\\
 Campo Boario, 64100 Teramo, Italy.\\
\textsuperscript{d}ISI Foundation\\
 Via Alassio 11/c, 10126 Torino, Italy.\\
}
\end{flushleft}

\newpage
 
 \section{Seasonal pattern in cattle trade network}
\begin{figure}[htbp]
\begin{center}
 \includegraphics[width=8cm]{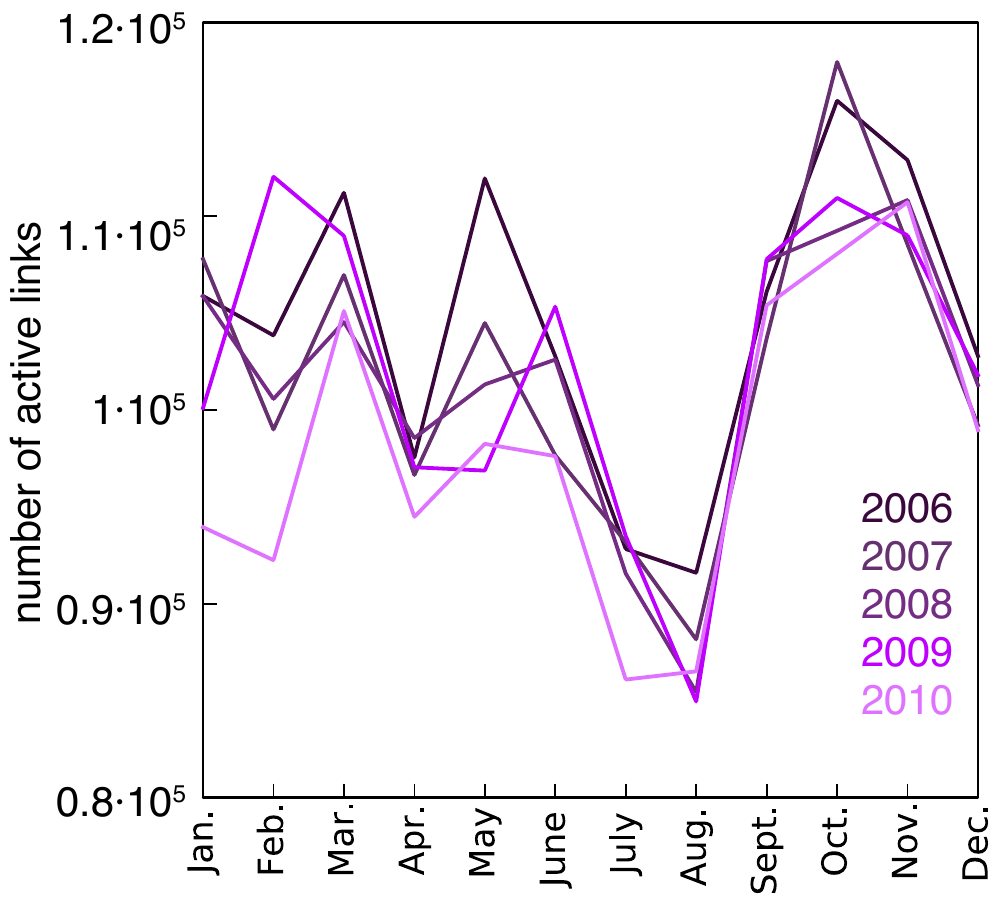}
\end{center}
\caption{
Number of active links per month in cattle trade network. Different colors pertain to different years, in range $(2006-2010)$.
}
\label{fig:seasonality}
\end{figure}

Fig.~\ref{fig:seasonality} shows the number of active links per month in the cattle trade network. A seasonal pattern is clearly visible: the activity drops during summer months, and peaks during fall.  The activity pattern is quite similar from one year to the other. 
\section{Loyalty's properties} 
%
\subsection{Allowed values}
In the following we provide an analytical reasoning on the allowed values for the loyalty. $\theta$ between configurations $c$ and $c+1$ can be rewritten as
\begin{equation}
 \theta  = \frac{\alpha}{A-\alpha},
\end{equation}
where $\alpha\in\mathbb{N}$ is the number of neighbors retained from $c$ to $c+1$, and $A=k_c+k_{c+1}\in\mathbb{N}$ is the sum of the node's degrees.
Clearly, every pair of $\alpha',A'$ for which $\exists q\in\mathbb{N}$ such that $\alpha'=q\alpha$ and $A'=qA$, will give the same $\theta$.
Therefore, in order to compute all the possible values of $\theta$, we must restrict ourselves to $\alpha,A$ coprimes: $(\alpha,A)=1$.
Moreover, since $\theta$ cannot be higher than $1$, we have to impose one further constraint: $\alpha < A/2$. All divisions are to be intended as integer divisions.

For zero loyalty, we have $\theta=0\Leftrightarrow \alpha=0$, for every positive $A$.
If $\theta>0$, we  need to count the number of possible values $\alpha$, given the constraints discussed above, and given a value for $A$ which is fixed by the node's degrees.
For $A\geq 3$, there are $\varphi(A)/2$ coprimes of $A$ and smaller than or equal to $A/2$, as it can be inferred by basic properties of the Euler's totient function $\varphi$.
\begin{equation}
 n(A) = 
 \left\{
 \begin{array}{lr}
 0 & \mbox{ if }A=1 \\
 1 & \mbox{ if }A=2 \\
 \varphi(A)/2 & \mbox{ if }A\geq 3 \\
 \end{array}
 \right. ,
\end{equation}
where $n(A)$ counts the number of nonzero allowed values for $\theta$, given a fixed $A$.
In order to compute the total number of allowed $\theta$ values in an entire network, we now let $A$ run from 1 to a certain $A_{max}$, which is of the order of twice the highest degree:
\begin{equation}
 {\mathcal N}(A_{max}) = 1 + \sum_{A=1}^{ A_{max} }n(A) = 2 + \frac{1}{2} \sum_{A=3}^{ A_{max} }\varphi(A).
\end{equation}
The unity added to the sum takes into account the value $\theta=0$. In order to better understand the behavior of ${\mathcal N}(A_{max})$ we can use Walfisz approximation for large $A_{max}$, and assume $A_{max} \approx 2k_{max}$ to get
\begin{equation}
 {\mathcal N}(k_{max}) = 1 + \frac{6}{\pi^2} k_{max}^2 + {\mathcal O}\left[  k_{max} \left( \log k_{max}  \right)^{2/3} \left( \log\log k_{max}  \right)^{4/3}   \right].
\end{equation}
This means that the sexual contact network has $\sim 10^4$ allowed values, and the cattle trade network has $\sim 10^8$ allowed values. Such large number of allowed values in the interval $\left[0,1\right]$ justifies our approximation of treating $\theta$ as a continuous variable.
%
%
\subsection{Temporal stability of the loyalty distribution in cattle and sexual contact networks}
\begin{figure}[htbp]
\begin{center}
 \includegraphics[width=12cm]{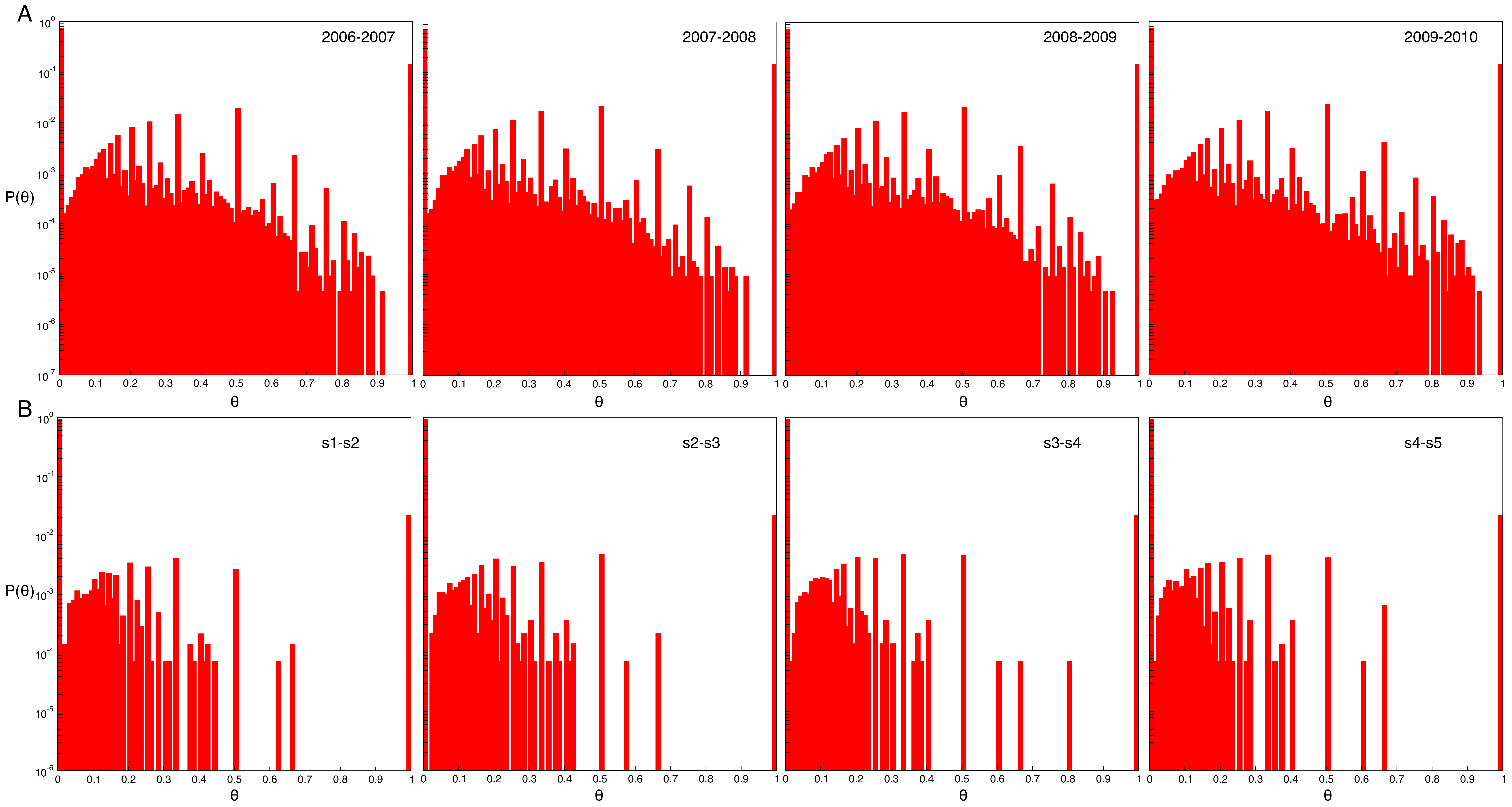}
\end{center}
\caption{
Loyalty distributions for different configurations. ({\itshape A}): distributions for cattle network, over the considered time period. ({\itshape B}): distributions for sexual contacts network.}
\label{fig:ptheta_sequence}
\end{figure}

Fig.~\ref{fig:ptheta_sequence} shows the loyalty distributions in all configuration pairs included in the two datasets under study (top,   cattle trade network; bottom,  sexual contact network). In both networks, distributions are stable in time.

\subsection{Correlation between loyalty and degree}
\begin{figure}[htbp]
\begin{center}
 \includegraphics[width=12cm]{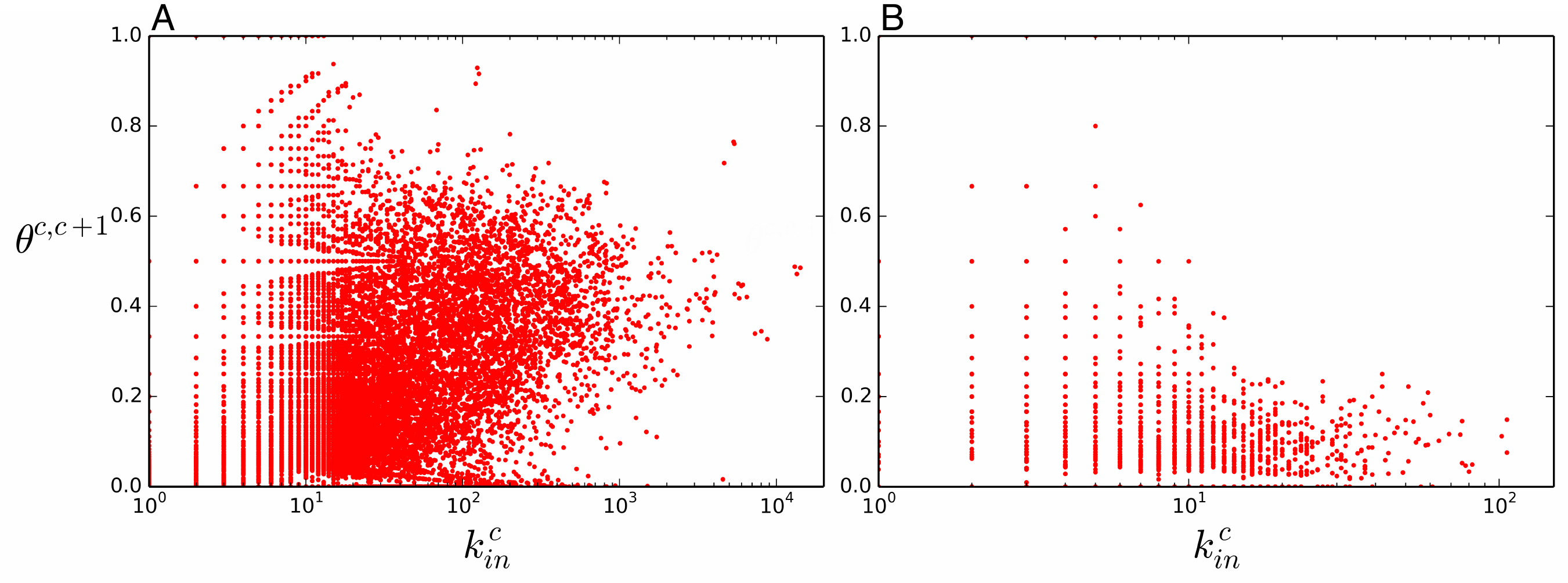}
\end{center}
\caption{
Scatter plots showing degree at configuration $c$ vs loyalty between configurations $c,c+1$. Each point represents a node. ({\itshape A}): cattle network. ({\itshape B}): sexual contacts.
}
\label{fig:scatter_kt}
\end{figure}

Degree and loyalty, while not being independent variables, are nonetheless not trivially correlated. Fig.~\ref{fig:scatter_kt} shows the scatter plots between the degree of a node in configuration $c$ and its loyalty for the pair of configurations $c,c+1$, for both networks. For each value of $k$, $\theta$ is found to range over a wide interval. This is clearly visible up to $k\approx 10^2$ for the cattle trade network, and $k\approx 10$ for the sexual contact network. Higher degree nodes are much less frequent, so the statistics becomes poorer and the heterogeneity in $\theta$ decreases as $k$ increases. 
Pearson correlation coefficients are found to be low for both networks ($0.04$ for the cattle trade network and $0.15$ for the sexual contact network), consistently with the observed large variations. They are however significantly larger than the coefficients \aggiu{of the null model: }\espu{obtained by reshuffling $\theta$ values (}95\% confidence interval of $(-0.002,0.002)$ and $(-0.006,0.007)$, for the cattle trade network and the sexual contact network, respectively\espu{), pointing}\aggiu{. This points} to a positive, albeit weak, correlation between degree and loyalty. \aggiu{The confidence intervals for the null model are obtained by randomly shuffling several times the sequence of $\theta$'s, in order to highlight any spurious correlation with the degree sequence.}

\section{\aggiu{Loyalty and other similarity measures}}
\begin{figure}[htbp]
\begin{center}
 \includegraphics[width=15cm]{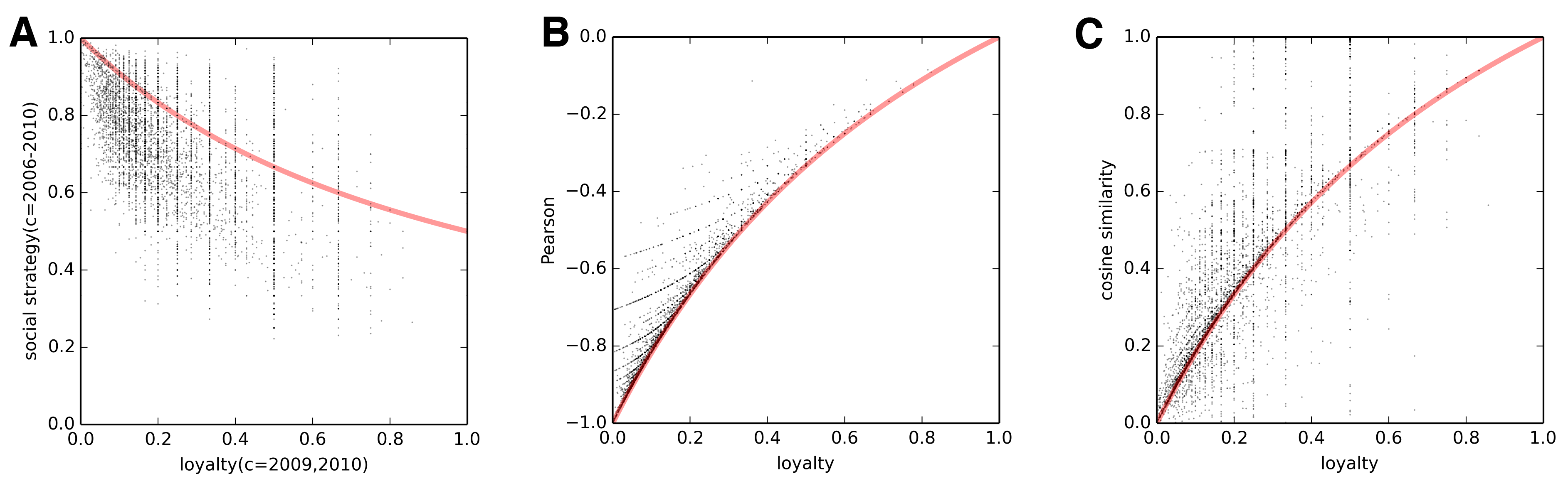}
\end{center}
\caption{\aggiu{
Cattle network: correlation between loyalty and other neighbor similarity measures. {\itshape (A)}: scatter plot showing social strategy ($\gamma$) computed from $2006$ to $2010$ vs loyalty between $2009,2010$. Each point represents a node. The red line represents $\gamma^{1,2}_i = \frac{1}{ 1+\theta^i_{1,2} }$; Pearson correlation is $-0.59$. {\itshape (B)}: Pearson ($\xi$) vs loyalty. The red line represents $\langle \xi^{c,c+1} \rangle$. {\itshape (C)}: cosine similarity $\zeta$ vs loyalty. The red line represents $\langle \zeta^{c,c+1} \rangle$.
}}
\label{fig:scatter_altremisure}
\end{figure}

\aggiu{
We analyze here the relationship between loyalty and other possible measures of similarity of the neighbor structure of a node across time. Firstly we consider a measure introduced as {\itshape social strategy} in [SM-1]. In our context, if we call $\tilde{k}^{1,c}_i$ the (in-)degree of node $i$ in the network resulting from the aggregation of snapshots $1$ to $c$, then $i's$ social strategy in those configurations will be computed as $\gamma^{1,c}_i = \tilde{k}^{1,c}_i / \left(\sum_{c'} k^{c'}_i\right)$. $k^{c'}_i$ is as usual the (in-)degree of $i$ in configuration $c'$.  This definition is the same as in \cite{miritello2013}, except for a normalizing factor $c$. We make this choice in order to make the comparison with $\theta$ more straightforward. The most important qualitative difference between loyalty and social strategy is that the former is always computed between a pair consecutive snapshots, while the latter typically describes an average behavior computed on several configurations (from $1$ to $c$ in our notation). Indeed only in the trivial case of $\gamma$ computed on just two snapshots, loyalty and strategy are univocally related: $\gamma^{1,2}_i = \frac{1}{ 1+\theta^i_{1,2} }$. In general, $\gamma^{1,c}_i$ will be a non trivial combination of all the consecutive loyalties $\theta^{1,2}_i,\theta^{2,3}_i\cdots\theta^{c-1,c}_i$ and degrees. Fig.~\ref{fig:scatter_altremisure}A shows the correlation between social strategy in cattle network, computed from $2006$ to $2010$, and loyalty between $2009,2010$.
}

\aggiu{
 We now consider a measure of neighbor similarity derived from Pearson correlation coefficient. This measure is analogous to what is called {\itshape adjacency correlation} in [SM-2]. For each node we build two vectors, $v_i^c,v_i^{c+1}$, of dimension $|\mca{V_i^{c}}\cup\mca{V_i^{c+1}}|$, i.e. these vectors will contain an entry for each vector that is neighbor of $i$ in at least one of the two configurations. $v_i^c$ has entries equal to $1$ for nodes that are in $\mca{V_i^{c}}$, and zero otherwise, and the same for $v_i^{c+1}$. We than consider the Pearson correlation coefficient between the two vectors,  $\xi^{c,c+1}$. This can be directly related to the loyalty $\theta_i^{c,c+1}$ and the degrees of the node in the two configuration $k^{c}_i$ and $k^{c+1}_i$ through the formula
}
\begin{equation}
\aggiu{
 \xi^{c,c+1} = - \frac{k^{c}+k^{c+1}}{\sqrt{k^{c}k^{c+1}}} \frac{1}{1+\theta^{c,c+1}} \sqrt{  \left(1+\theta^{c,c+1} \right)^2 \frac{k^{c}k^{c+1}}{ \left(k^{c}+k^{c+1}\right)^2 } - \theta^{c,c+1}  }
 }
\end{equation}
\aggiu{
In the above equation we have omitted the subscript $i$: $ \xi^{c,c+1}= \rho^{c,c+1}_i$, $k^{c}=k^{c}_i$ and $\theta^{c,c+1}=\theta^{c,c+1}_i$.  Fig.~\ref{fig:scatter_altremisure}B shows the scatter plot  $\xi^{c,c+1}$ versus $\theta^{c,c+1}$. We see that, due to the definition of vectors $v^{c}$, $\xi\in[-1,0]$.
This formula can be simplified if we need just an average behavior: assuming $k^{c}=k^{c+1}=k$, where $k$ is the average connectivity, the formula reduces to $\langle \xi^{c,c+1} \rangle = -(1-\theta^{c,c+1})/(1+\theta^{c,c+1})$. From this we get that $\theta=0$ (no memory) corresponds to $\xi=-1$, while $\theta=1$ (perfect memory) corresponds to $\xi=0$.
}

\aggiu{
Finally, we analyze an application of cosine similarity. For each node vectors $v_i^c,v_i^{c+1}$ are built as before. Then cosine similarity between those vectors is defined as $\zeta = v_i^c\cdot v_i^{c+1} / \left( |v_i^c| |v_i^{c+1}|\right)$. It can be shown that, like $\xi$, $\zeta$ can be written in terms of degree and loyalty:
}
\begin{equation}
\aggiu{
 \zeta^{c,c+1} = \frac{\theta}{1+\theta} \frac{ k^{c}+k^{c+1} }{ \sqrt{k^{c}k^{c+1}}  }
 }
\end{equation}
\aggiu{
The average behavior this time is $\langle \zeta^{c,c+1} \rangle = 2 \theta^{c,c+1}/(1+\theta^{c,c+1})$ (see scatter plot in Fig.~\ref{fig:scatter_altremisure}C).
}

\aggiu{
In conclusion, social strategy, being computed on a sequence of more than two configurations, represents a qualitatively different measure with respect to loyalty, albeit the two measures being correlated (see Fig.~\ref{fig:scatter_altremisure}A). On the other hand, both Pearson $\xi$ and cosine similarity $\zeta$ can be completely determined in terms of degree and loyalty. Moreover, the mean trend is well modeled by the averaged version of these measure, which discounts degree (see Fig.~\ref{fig:scatter_altremisure}). 
}

%
%
\section{Modeling infection potentials}
Infection potentials $\pi_D$ and $\pi_L$ are modeled with a sum of an exponential distribution, to account for the behavior at $\pi\simeq 0$, and a Landau distribution, to mimic the particular asymmetry around the peak. The exact formulation is the following:
\begin{equation}
 f \left(x;\mu,\sigma,r,q\right) \propto \exp\left(-qx\right) + r \int_0^\infty dt \, \sin (2t) \exp\left[ -t \frac{x-\mu}{\sigma} -\frac{2}{\pi} t \log t   \right].
\end{equation}
There are four free parameters: one for the exponential distribution, two for the Landau distribution, and one driving the relative importance of one function with respect to the other. An overall scaling coefficient is fixed by normalization.
\section{Robustness of the risk assessment procedure in varying parameters and assumptions}

\subsection{Threshold $\epsilon$}
\begin{figure}[htbp]
\begin{center}
 \includegraphics[width=12cm]{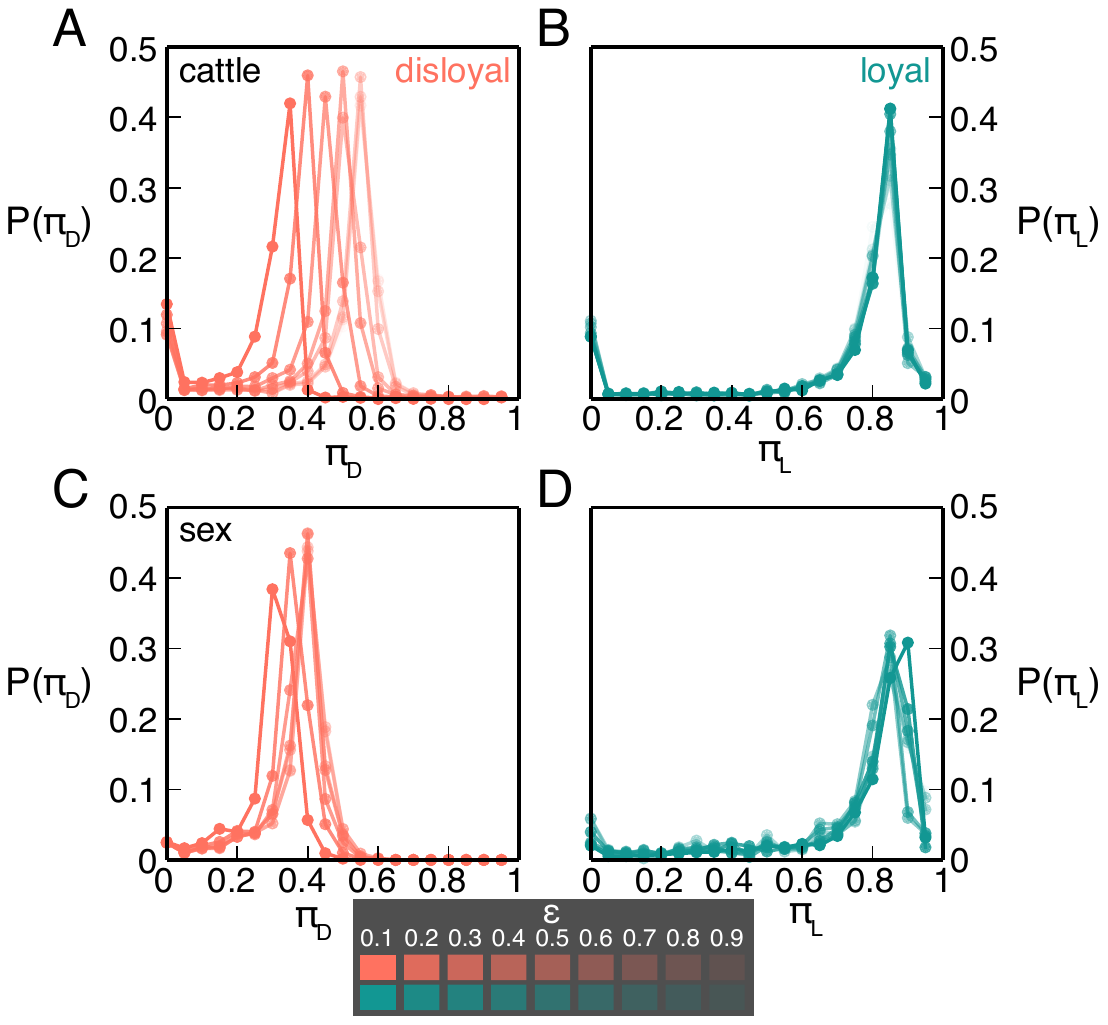}
\end{center}
\caption{
Behavior of infection potentials $\pi_D$ and $\pi_L$ as $\epsilon$ varies. ({\itshape A}),({\itshape C}): $\pi_D$ curves. ({\itshape B}),({\itshape D}): $\pi_L$ curves. ({\itshape A}),({\itshape B}): cattle network. ({\itshape C}),({\itshape D}): sexual contacts.
}
\label{fig:epsilon_sensit}
\end{figure}

In the following we examine the behavior of the infection potentials $\pi_D$ and $\pi_L$  in varying the value of the threshold.
Fig.~\ref{fig:epsilon_sensit} shows that in the cattle trade network the peak position of $\pi_D$ increases with $\epsilon$, from 0.3 to 0.6. Such behavior is present in the sexual contact network too, albeit less evident (from 0.3 to 0.5).
Unlike $\pi_D$, $\pi_L$ distributions remain stable as $\epsilon$ varies.
As a result, the probability of a loyal node being infected ($\pi_L$) does not depend on the choice of $\epsilon$. The choice of threshold $\epsilon=0.1$ thus allows to maximize the distance between $\pi_D$ and $\pi_L$ distribution while preserving enough statistics for the loyal nodes.

It is important to note that the value of $\epsilon$ also affects  the transition probabilities $T_{DD},T_{LL}$ in their functional dependence on the degree (Figure 3C,D of the main text). For each threshold value, such dependence needs therefore to be 
assessed through a fitting, to be used for the prediction of the loyalty values in the unknown network configuration.

\begin{figure}[htbp]
\begin{center}
 \includegraphics[width=12cm]{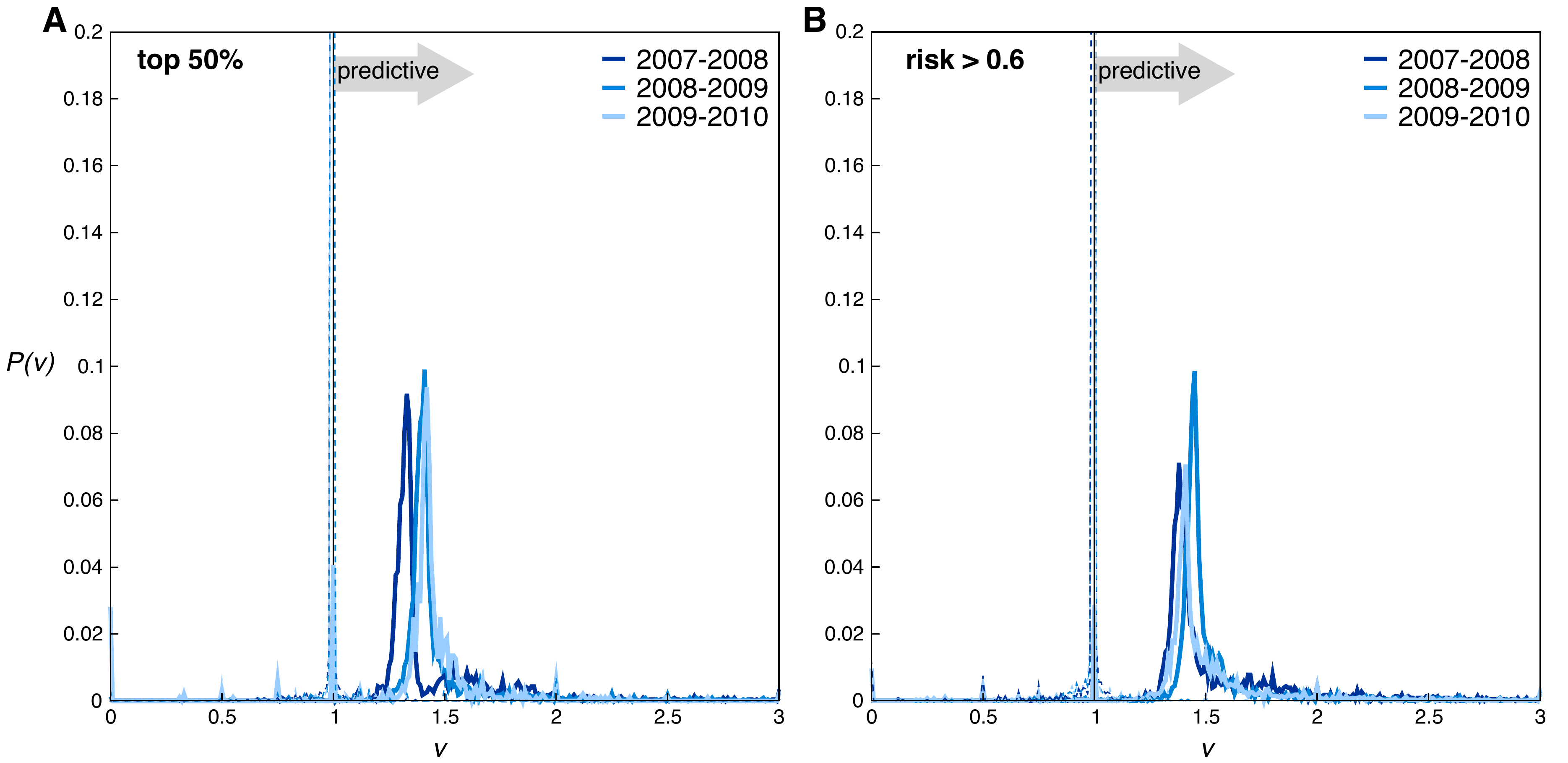}
\end{center}
\caption{
Risk ratio ($\nu$) distribution for cattle network. ({\itshape A}): ${\mathcal I}^c_{s,h}$ as set of top $50\%$ highest ranking nodes. ({\itshape B}): ${\mathcal I}^c_{s,h}$ as set of nodes with $\rho>0.6$.
}
\label{fig:rrcattle}
\end{figure}
\subsection{Definitions for the risk ratio $\nu$}
In the main paper the risk ratio $\nu$ is computed considering the set ${\mathcal I}^c_{s,h}$ of the top $25\%$ highest ranking nodes. Here we explore two different ways of defining this quantity:
\begin{itemize}
 \item ${\mathcal I}^c_{s,h}$ as the set of the top $50\%$ highest ranking nodes (Fig.~\ref{fig:rrcattle}A);
 \item ${\mathcal I}^c_{s,h}$ as the set of nodes with epidemic risk $\rho>0.6$ (Fig.~\ref{fig:rrcattle}B).
\end{itemize}
Results are reported in Fig.~\ref{fig:rrcattle} showing the invariance of the observed $\nu$ results on this arbitrary choice.

\subsection{Definition of the early stage of an epidemic}

\begin{figure}[h!]
\begin{center}
 \includegraphics[width=12cm]{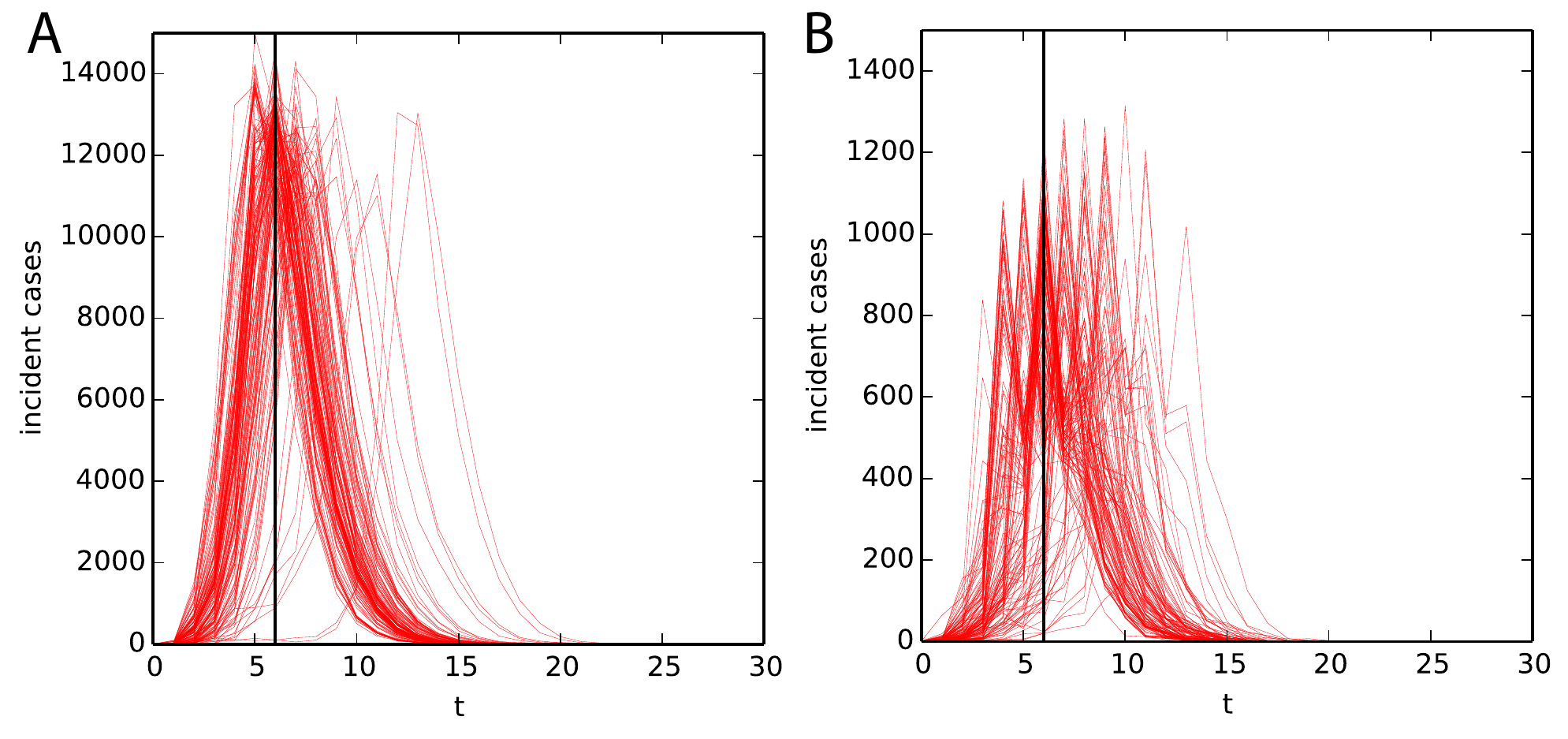}
\end{center}
\caption{\aggiu{
Simulated incidence curves obtained by changing seeding node and network configuration for the cattle trade network ({\itshape A}), and the sexual contacts network ({\itshape B}). Black line indicates $\tau=6$.
}
}
\label{fig:incidence}
\end{figure}

In the main paper we consider an initial stage of the epidemic  up to $\tau=6$.
\aggiu{
This choice being arbitrary, it is informed by the simulated time behavior of the incidence curves (see Fig.~\ref{fig:incidence})
and the aim to focus on the initial stage of the epidemic.}

\begin{figure}[htbp]
\begin{center}
 \includegraphics[width=12cm]{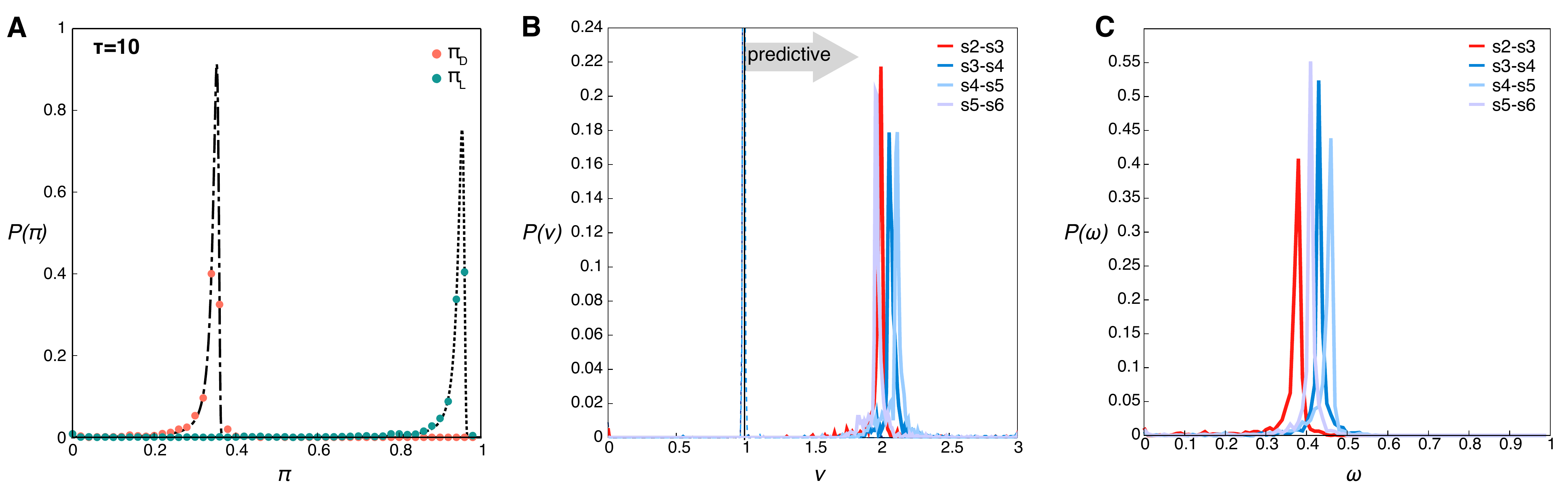}
\end{center}
\caption{
Invasion stage of the outbreak up to $\tau=10$ for sexual contacts network: distribution of the infection potentials ({\itshape A}), the risk ratio $\nu$ ({\itshape B})and the predictive power $\omega$ ({\itshape C}).
}
\label{fig:var_T}
\end{figure}

\espu{If we}\aggiu{We also} tested a longer initial stage ($\tau=10$) \aggiu{for the sexual contacts network, to assess the impact of this variation on the obtained results.}\espu{,} We obtain distributions of the infection potential, of the relative risk ratio, and of the predictive power showing sharper peaks, however with unchanged peak positions (Fig.~\ref{fig:var_T} for the sexual contact network). Peaks are expected to be sharper, because with $\tau=10$ a larger fraction of the network is reached by the outbreak. The fact that peak positions do not change, however, reveals that we are able to provide accurate epidemic risks already at the earlier phase of the epidemic  ($\tau=6$), when such information is mostly needed.
%
\subsection{Aggregation time window \espu{for the sexual contact network}}

The choice of yearly aggregation time in the case of  the cattle trade network is informed by its annual seasonal dynamics; the six-months aggregating window for the sexual contact network is instead arbitrary. Here we explore other aggregating windows for both networks to explore the impact they may have on the obtained results.

\begin{figure}[htbp]
\begin{center}
 \includegraphics[width=12cm]{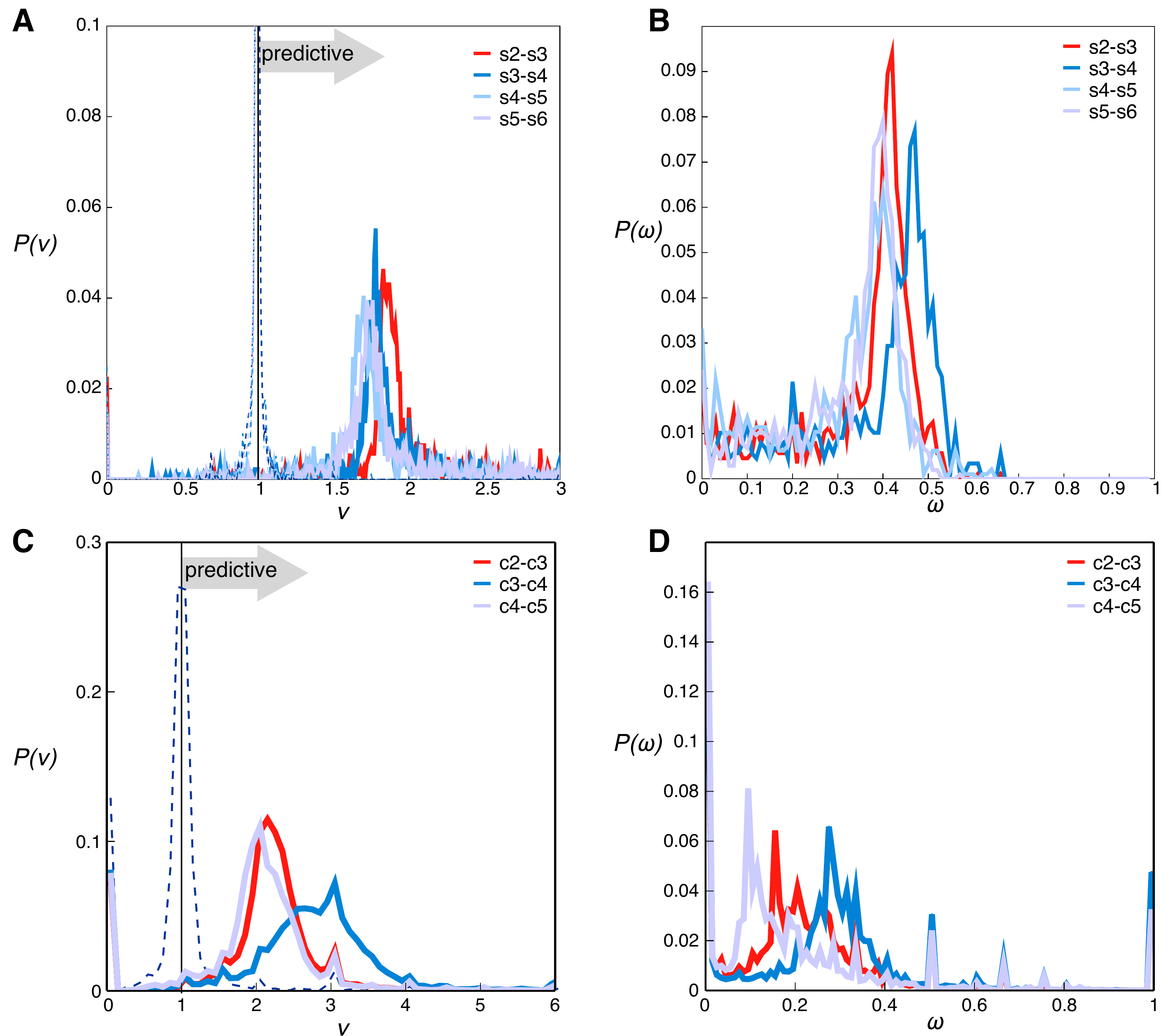}
\end{center}
\caption{
\aggiu{
Exploration of different aggregating windows (cattle: 4-month, sex: 3-month). Distributions of risk ratio $\nu$ in cattle ({\itshape A}) and sexual contacts ({\itshape C}). Distribution of predictive power $\omega $ in cattle ({\itshape B}) and sexual contacts ({\itshape D}).
}
\espu{3-months aggregating windows for sexual contacts network: distribution of the infection potentials ({\itshape A}), the risk ratio $\nu$ ({\itshape B}) and the predictive power $\omega$ ({\itshape C}). }
}
\label{fig:var_m}
\end{figure}

We consider configurations for the sexual contact network consisting of 3-months aggregation. \espu{Fig.~\ref{fig:var_m}A is the analogous of Fig.~3B in the main paper, and shows the shape of the infection potential distributions for this particular choice of aggregation. We observe a similar behavior in the two pictures. The only qualitative difference is that when aggregation is performed with 3-months windows the distributions appear to be broader and noisier with respect to 6-months ones, probably due to the increased sparseness of the network configurations.}
When calculating the risk ratio and the predictive power \aggiu{(Fig.~\ref{fig:var_m}B,D)}, we find distributions similar to the ones reported in the main text, with unchanged peak positions. \espu{Analogously to what happens for the infection potential, t}\aggiu{T}he distributions \aggiu{however} appear to be noisier, especially as far as $\omega$ is concerned, \aggiu{likely induced by the increased sparseness of the network configurations}.

\aggiu{
We also try a different aggregation time for cattle network: 4-month windows. Risk ratio and predictive power distributions are presented in Fig.~\ref{fig:var_m}A,C. We observe that $\omega$ is on average quite low: this is likely due to the fact that aggregation windows shorter than one year fail to take into account the seasonal patterns, thus decreasing system memory.
}

\section{Memory driven model: analytical understandings}
\subsection{Amount of memory}
In the following we analytically quantify the amount of memory in the memory driven model as the probability $f_{c,c+1}$
that a link present in configuration $c$ is also present in configuration $c+1$. This can be expressed as:
\begin{equation}
 f_{c,c+1} = (1-d)\left[  p_\alpha + \frac{1}{N}\frac{b(1-d)}{b+d}\frac{\zeta(\gamma-1)}{\zeta(\gamma)}  \right],
 \label{eq:memoria}
\end{equation}
where the first term, $(1-d)p_\alpha$, is the probability of remaining active and at the same time keeping a particular neighbor. The second term is the probability of not keeping a neighbor but recovering it with one of the new stubs. $\zeta$ is the Riemann $\zeta$-function.
$f_{c,c+1}$ can indeed be interpreted as the system memory, as it is a good estimator of the fraction of links that survive from one configuration to the following.

The second term in Eq.~\ref{eq:memoria} is suppressed by $1/N$ and can be disregarded in our case given the large size of the networks ($N=10^4$). $f_{c,c+1} \approx (1-d)p_\alpha$ therefore provides a first order approximation that correctly matches the numerical results (see Fig.~\ref{fig:model_anal}A for the comparison).

\subsection{Probability associated to zero loyalty}
The probability of a node with in-degree $h_c$ having zero loyalty ($\theta_{c,c+1}=0$) can be computed analytically as
\begin{equation}
 P(\theta_{c,c+1}=0|k_c) = d+(1-d)\left(1-p_\alpha\right)^{k_c}.
\end{equation}
\begin{figure}[h!]
\begin{center}
 \includegraphics[width=12cm]{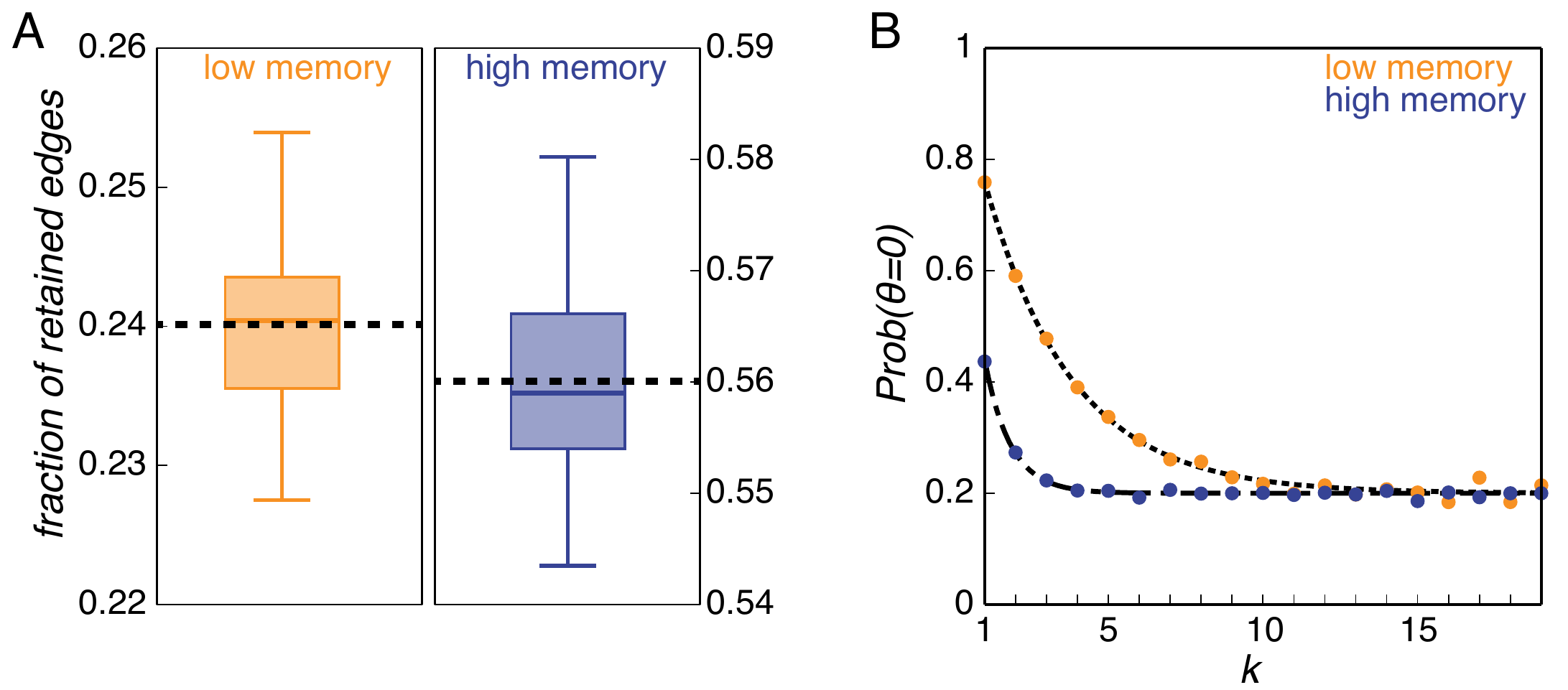}
\end{center}
\caption{
Characterization of the memory driven dynamical model.
({\itshape A}): the memory of the system, in terms of the fraction of edges retained from one configuration to the following. Boxplots represent median and quartile positions. The distributions are computed over 50 realizations of the model. Dashed lines represent the theoretical prediction. $p_\alpha=0.3,0.7$ for low and high memory, respectively.
({\itshape B}): probability for a node with a given in-degree $k$ to be completely disloyal $(\theta=0)$ between two following snapshots. Points represent numerical simulations, while lines show the theoretical estimates.
}
\label{fig:model_anal}
\end{figure}
In Fig.~\ref{fig:model_anal}B we check this result against numerical simulations.
\section{Memory driven model: additional properties}
\begin{figure}[htbp]
\begin{center}
 \includegraphics[width=12cm]{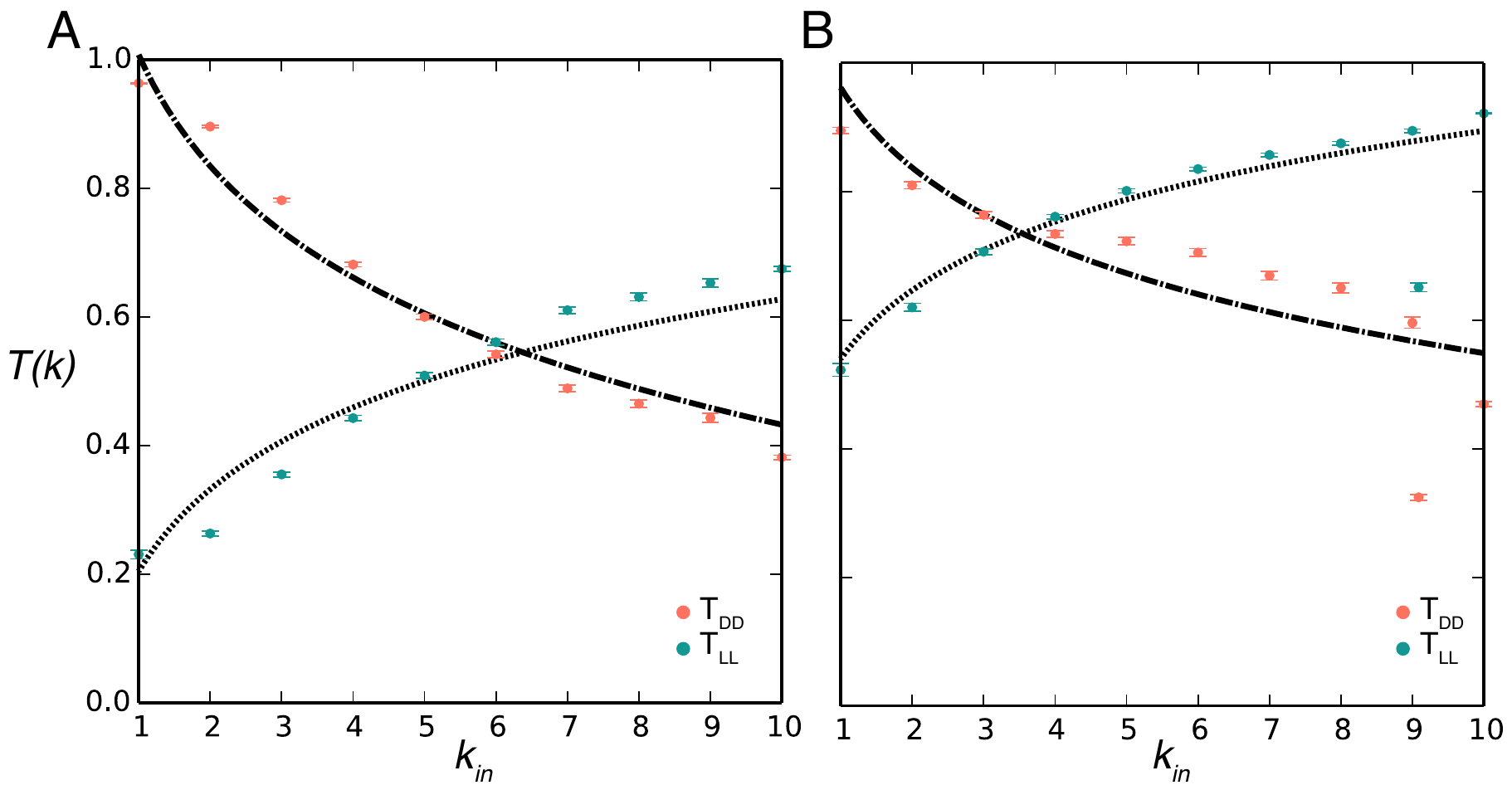}
\end{center}
\caption{
Memory driven model: loyalty transition probabilities between loyal statuses ($T_{LL}(k)$, green) and disloyal statuses ($T_{DD}(k)$, orange) as functions of the degree $k_{in}$ of the node.
({\itshape A}):  low memory model ($p_\alpha=0.3$), ({\itshape B}):  high memory model ($p_\alpha=0.7$).
Dashed lines represent the logarithmic models: $T_{DD}(k)=1.01-0.25\log k$, and $T_{LL}(k)=0.20+0.18\log k$ for the low memory; $T_{DD}(k)=0.96-0.17\log k$, and $T_{LL}(k)=0.53+0.15\log k$ for high memory.
 Error bars represent the deviation $\pm \left\{T(k)\left[1-T(k)\right]/N_k\right\}^{1/2}$, where $N_k$ is the number of nodes with degree $k$ used to compute $T(k)$. Last value for $k$: $k=10$ includes all nodes with degree equal or higher.}
\label{fig:model_trans}
\end{figure}

In the main paper the transitions probabilities between loyalty statuses are shown only for the real networks (main paper Fig.~3C and 3D). Here we present them for the memory driven model. Fig.~\ref{fig:model_trans} reports these probabilities in case of low and high memory, along with the modeling functions.

\begin{figure}[htbp]
\begin{center}
 \includegraphics[width=12cm]{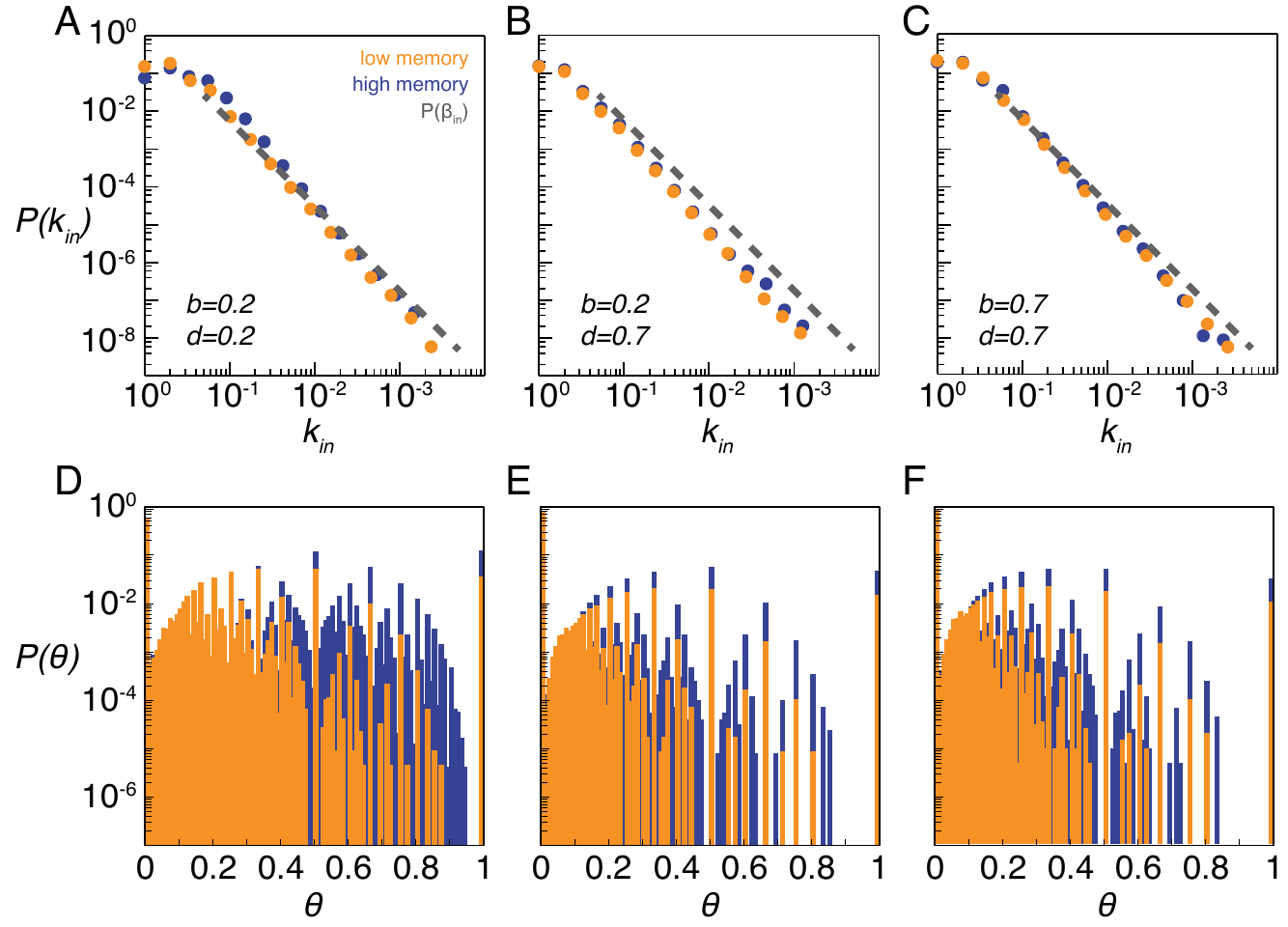}
\end{center}
\caption{
Memory driven model: degree and loyalty distribution when changing the probabilities of becoming active or inactive. ({\itshape A}),({\itshape B}),({\itshape C}): in-degree distributions when $(b,d)=(0.2,0.2),(0.2,0.7),(0.7,0.7)$, respectively. ({\itshape C}),({\itshape D}),({\itshape E}): loyalty distributions for the same parameter configurations.
}
\label{fig:param_explore}
\end{figure}

In addition, we explore different values of the model parameters and discuss the changes in the network properties.  In particular, we explore different values for the probability of becoming active ($b$) or inactive ($d$), other than the choice used in main paper ($b=0.7$, $d=0.2$).
Fig.~\ref{fig:param_explore}A, \ref{fig:param_explore}B, \ref{fig:param_explore}C are the equivalent of main paper Fig.~5A, and show the in-degree distribution for different values of $b,d$ in the set $\left\{0.2,0.7\right\}$. $P(k_{in})$ is very robust when changing these parameters, and in all cases follows the slope of the $\beta_{in}$ distribution.
 Fig.~\ref{fig:param_explore}D, \ref{fig:param_explore}E, \ref{fig:param_explore}F are the equivalent of main paper Fig.~5B, and show the loyalty distributions. We observe that the overall shape is insensitive to parameters change. There is however, a tendency to have higher $\theta$ values for low $b,d$. This is to be expected, since higher probabilities of going from active to inactive and vice versa mean larger turnover, which leads to lower memory and therefore lower overall loyalty.

\begin{figure}[htbp]
\begin{center}
 \includegraphics[width=12cm]{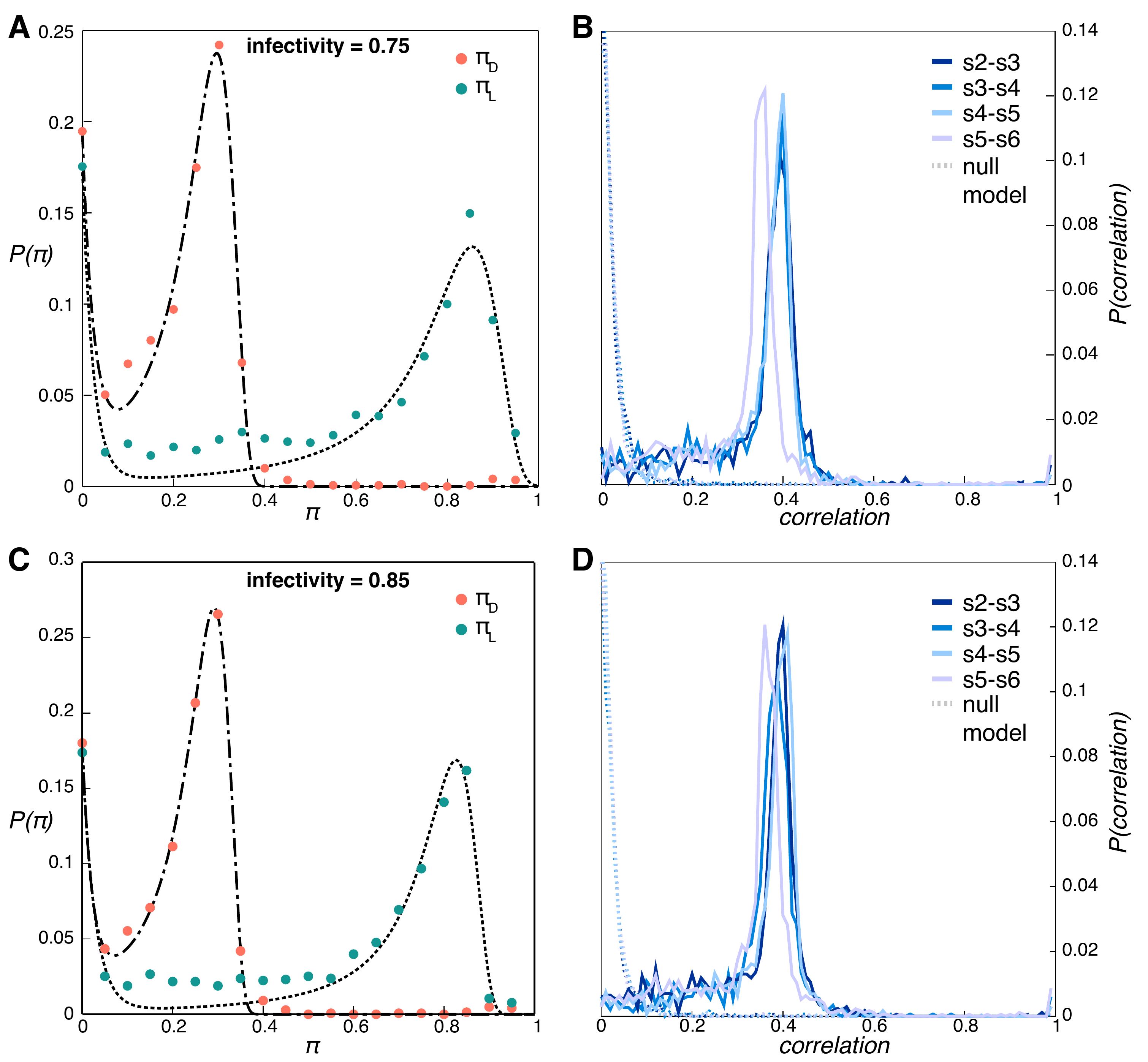}
\end{center}
\caption{
Applying the methodology to sexual contact networks using a stochastic epidemic model. ({\itshape A}),({\itshape C}): infection potentials for infectivity $0.75$ and $0.85$, respectively.  ({\itshape B}),({\itshape D}): distribution of the Pearson correlation coefficient between the computed epidemic risks and the probability of actually being infected, for infectivity $0.75$ and $0.85$, respectively. Dashed lines show distributions from the null model.}
\label{fig:stoch}
\end{figure}
\section{Validation in the stochastic case} 
\label{sec:validation_stoch}
We repeat the analysis reported in the main text by considering a stochastic Susceptible-Infectious approach. Given the same initial conditions, we perform $r$ different stochastic runs, each leading to potentially different outcomes. For each node $i$, we compute the fraction $f_i(s)$ of runs that node $i$ is infected from epidemics starting from seed $s$ within time step $\tau$. For validation, we need to compare the list $\left\{\rho_i\right\}(s)$ of the node epidemic risks computed with our methodology with the list $\left\{f_i\right\}(s)$ of the probabilities of actually getting infected. If our estimated risks are reliable, then the two lists need to be correlated, as a higher risk should correspond to a higher probability to get infected. In order to evaluate this, we compute the Pearson correlation coefficients between $\left\{\rho_i\right\}(s)$ and $\left\{f_i\right\}(s)$, for each possible seed $s$.  The list of these coefficients can then be summarized in a distribution. Fig.~\ref{fig:stoch}B and \ref{fig:stoch}D show such distributions for the sexual contact network for two different values of the infection transmissibility ($0.75$ and $0.85$, respectively).
In order to check that the correlation coefficients are significantly different from zero, we compute the same distributions after reshuffling the epidemic risks (dashed lines in plots).
Fig.~\ref{fig:stoch}A and \ref{fig:stoch}C are the equivalent of Fig.~3B in main paper and show that the peak position of the infection potential does not change from the deterministic case. Noise and peak width, however, increase considerably, as well as the probability of having $\pi_D=0$, and this effect is more pronounced for lower infection transmissibilities.
\section{\aggiu{Cattle network: taking into account links weights}} 
\begin{figure}[htbp]
\begin{center}
 \includegraphics[width=12cm]{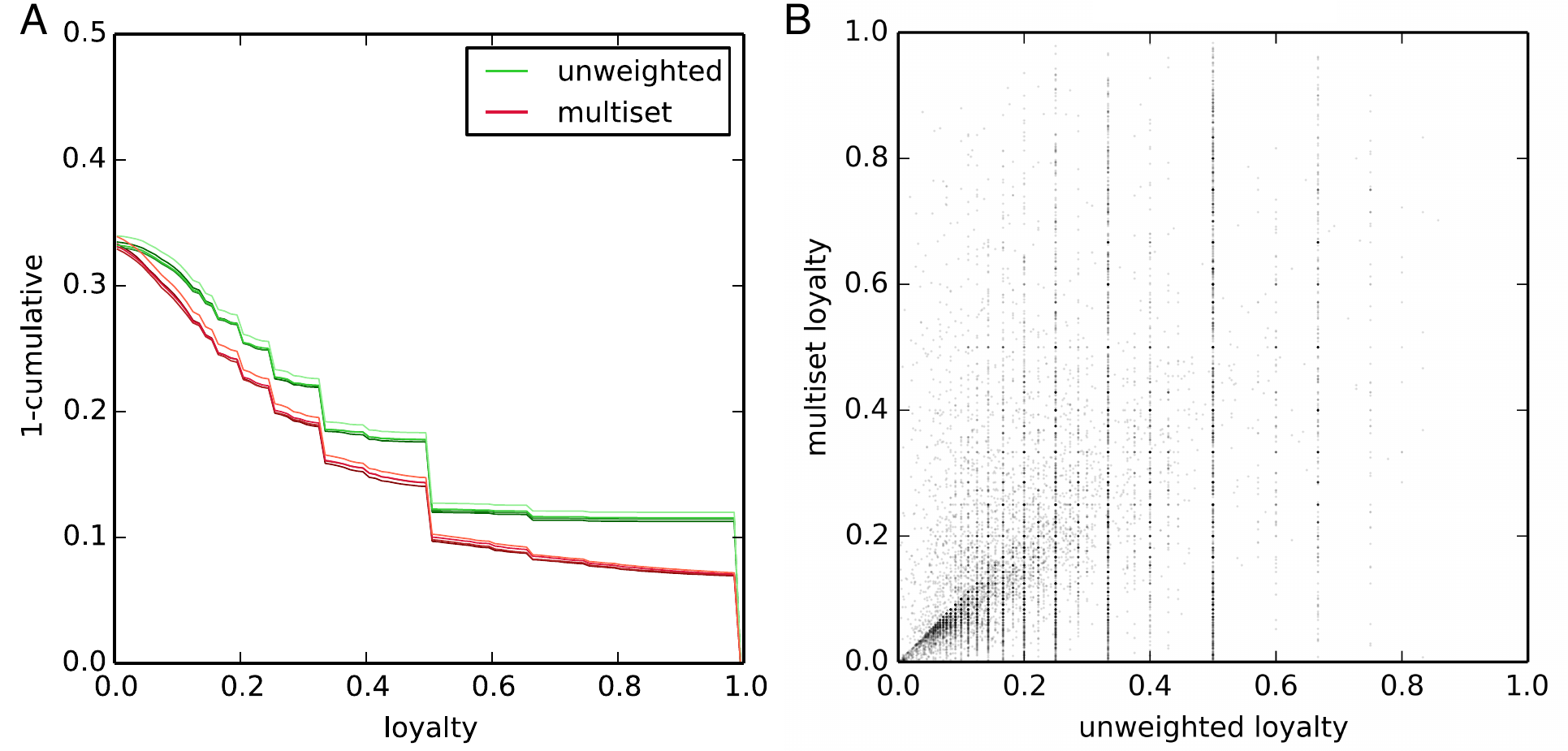}
\end{center}
\caption{
\aggiu{Weighted cattle network: extending the definition of loyalty. {\itshape A} shows the cumulative distributions for the unweigthed loyalty (green) and multiset loyalty (red). Different tones of colors refer to different network configurations. {\itshape B} Scatter plot correlating the unweighted loyalty and the multiset loyalty. Pearson correlation coefficient is $0.92$. }
}
\label{fig:loyalty_weight}
\end{figure}

\begin{figure}[htbp]
\begin{center}
 \includegraphics[width=12cm]{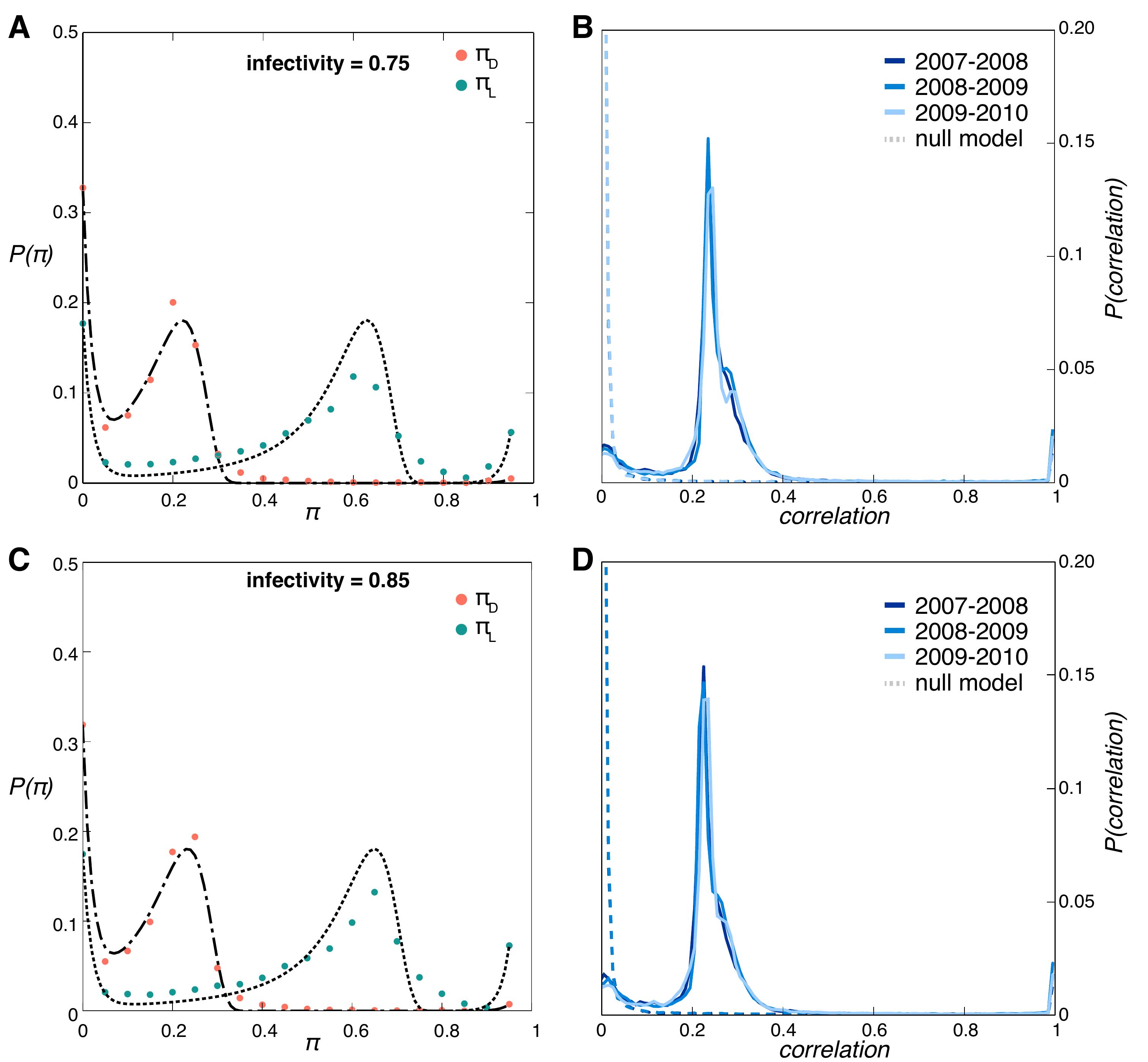}
\end{center}
\caption{
\aggiu{Weighted cattle network: risk prediction computation and validation. ({\itshape A}),({\itshape C}): infection potentials for single-animal infectivity $0.75$ and $0.85$, respectively.  ({\itshape B}),({\itshape D}): distribution of the Pearson correlation coefficient between the computed epidemic risks and the probability of actually being infected, for infectivity $0.75$ and $0.85$, respectively. Dashed lines show distributions from the null model. }
}
\label{fig:valid_weight}
\end{figure}

\aggiu{
Links in cattle network can be assigned a weight attribute in terms of the number of moved animals. These additional data can be included in the modeling of diseases spread, assuming that larger batches have a greater probability of carrying the disease from the source holding, to the destination. This feature is included in the disease model, by assuming a per-animal transmissibility $\lambda$. Then, given a movement of $w$ animals, the transmission probability along that link will be $\left[1-(1-\lambda)^w\right]$ (same approach as in SI of [SM-2012].
Loyalty needs to be generalized to the case of weighted network, too. The most straightforward generalization is obtained by considering the quantities in Eq.~(2) of main paper $\mca{V}_i^{c-1},\mca{V}_i^{c}$ as multisets (see, for instance, [SM-4]), where each neighbor appears as many times as the weight of the corresponding link. Then the {\it weighted loyalty} on the weighted network is defined, as before, by Eq.~(2) of main paper, using the definitions of multiset union and intersection: $ \mca{V}_i^{c-1} \cup \mca{V}_i^{c} = \sum_j \max\left( w^{c-1}_{ji}, w^c_{ji} \right)$ and $\mca{V}_i^{c-1} \cap \mca{V}_i^{c} = \sum_j \min\left( w^{c-1}_{ji}, w^c_{ji} \right)$, where $w^c_{ji}$ is the weight of the link $j-i$ in configuration $c$ (assuming $w=0$ if no such link is present). Other choices of similarity between sets of neighbors are possible, however this one is the most natural generalization, since it has a very similar distribution to the unweighted loyalty (Fig.~\ref{fig:loyalty_weight}A), and correlates well with it (Fig.~\ref{fig:loyalty_weight}B).
We now compute the infection potentials and then the epidemic risks, using this new loyalty. We validate the computed risks analogously to what we did in Sec.~\ref{sec:validation_stoch}. Results are presented in Fig.~\ref{fig:valid_weight}, showing the generalizability of our approach to the weighted case too.
}

\section{\aggiu{Assessing the robustness of risk based prediction with respect to simple predictors}}
\label{sec:multivar}
\aggiu{
We have shown that $\rho$ effectively represents the risk of being infected, as shown in the Validation section of main paper. We now show that $\rho$ is a significant improvement in prediction accuracy, with respect to simpler measures, like the degree of a node. From configurations $c-1,c$ of cattle network we compute the risk of being infected at $c+1$: $\rho_i = \rho_i^{c+1}(s)$, as in Eq.~(3) of main paper. For each node $i$ for which we can compute $\rho_i$ we then have the binary variable $outcome$ indicating if node $i$ is eventually hit by the epidemic in configuration $c+1$. We perform a multivariable logistic regression to check that $\rho$ is actually a predictor for $outcome$, adjusting the in-degree in configuration $c$: $k_i^{c}$. In particular, due to the high heterogeneity of $k$, we adjust for the $\log$ of the degree. Tab.~\ref{tab:multivar} shows the results of the performed regressions. As the crude odds ratios show, both $\rho$ and $k^c$, on their own, are meaningful predictor of infection in configuration $c+1$. We are however interested in assessing wether our risk is still a predictor, once the effect of knowing the degree is discounted for. The odds ratio for $\rho$ adjusted for degree is still significantly greater than one, meaning that even within nodes of the same degree, nodes at high risk are likelier to get infected. In other words, computing the risk (for which the knowledge of the degree of the node is needed) gives more predicting power than the sole knowledge of degree.
}
\begin{table}[htbp]
\centering
\begin{tabular}{rccc}
& \bf{crude OR} &\phantom{spa}& \bf{adjusted OR} \\
 \hline\hline
  \bf{log(degree)} & $2.88\;\;[2.87,2.89]$ &\phantom{spa}& $2.08\;\;[2.07,2.10]$\\
 \bf{risk}              & $4.82\;\;[4.78,4.86]$ &\phantom{spa}& $2.50\;\;[2.49,2.51]$\\
 \hline\hline
\end{tabular}
\caption{\aggiu{Odds ratios of being infected in configuration $c+1$, given degrees in $c$ and computed risks. Crude odds ratios refer to two separate univariate regressions; adjusted odds ratios are obtained through a single multivariate regression. $95\%$ confidence intervals are reported.} }
\label{tab:multivar}
\end{table}

\section{\aggiu{Application to human proximity networks}}
\aggiu{
The main difficulty in applying our methodology to physical proximity networks in human is that generally those networks are much smaller than the ones we have examined, that making it difficult to reach enough statistics to fit the form of infection potentials and transitions probability, and then perform the validation. We show here how we can overcome these impairments and apply successfully our strategy to a network of face-to-face proximity at a scientific conference, collected by the Sociopatterns group [SM-5]. This network records the interactions of $113$ nodes during a period of $2.5$ days. We split such networks in $30$ configurations (corresponding to hourly time steps), and use the first $29$ configurations to train our methodology, in order to give predictions on the $30th$. We use this large number of configurations in order to be able to build reliable empirical distributions for the infection potentials and the transition probabilities between loyalty statuses. Once risks are computed as usual, it is not possible, however, to perform the validation as we did for cattle, sexual contacts and memory driven models. This impossibility arises from the fact that the computed risk ratios are too few to build their distribution. In order to validate our methodology we therefore use the same technique implemented in Sec/~\ref{sec:multivar}: for every node, we compute the odds ratio of being infected in the last configuration, given the knowledge of degree and the computed risk. Results are reported in Tab.~\ref{tab:ht09}. Computer risks are strong predictors for infection, even after adjusting for degree. Moreover, unlike cattle network (see Tab.~\ref{tab:multivar}), degree alone is not a predictor. Predictive power $\omega$ is on average high: median $0.87$, with quartiles $Q_1:0.69,Q_2:0.97$.
}

\begin{table}[htbp]
\centering
\begin{tabular}{rccc}
& \bf{crude OR} &\phantom{spa}& \bf{adjusted OR} \\
 \hline\hline
  \bf{log(degree)} & $1.16\;\;[1.13,1.20]$ &\phantom{spa}& $0.95\;\;[0.89,1.02]$\\
 \bf{risk}              & $11.97\;\;[7.79,18.4]$ &\phantom{spa}& $22.34\;\;[7.90,63.3]$\\
 \hline\hline
\end{tabular}
\caption{\aggiu{Odds ratios of being infected in last configuration last, degree and computed risk. Crude odds ratios refer to two separate univariate regressions; adjusted odds ratios are obtained through a single multivariate regression. $95\%$ confidence intervals are reported.} }
\label{tab:ht09}
\end{table}

\begin{flushleft}
 {\sffamily\mdseries\upshape \bfseries\large SM References}\\
 \vspace{0.5cm}
 {[}SM-1] Miritello G, Lara R, Cebrian M, Moro E (2013) Limited communication capacity unveils strategies for human interaction. {\embib Sci Rep} 3, 1950. \\
 \vspace{0.2cm}
 {[}SM-2] Clauset A, Eagle N (2012) Persistence and periodicity in a dynamic proximity network, {\embib ArXiv}:1211.7343. \\
 \vspace{0.2cm}
 {[}SM-3] Bajardi P, Barrat A, Savini L, Colizza V (2012) Optimizing surveillance for livestock disease spreading through animal movements. {\embib J Roy Soc Int} (June, 2012). \\
 \vspace{0.2cm}
 {[}SM-4] Stanley R P (1997). Enumerative Combinatorics, Vols. 1 and 2., {\embib Cambridge University Press}. \\
 \vspace{0.2cm}
 {[}SM-5] Isella L, Stehl{\'e} J, Barrat A, Cattuto C, Pinton J-F, Van den Broek Wouter (2011) What's in a crowd? Analysis of face-to-face behavioral networks, {\embib Journal of Theoretical Biology} 271,1:166-180. \\
\end{flushleft} 

\end{document}